\documentclass[twocolumn,superscriptaddress, amsmath,amssymb,aps, pre,
showkeys,showpacs,nofootinbib] {revtex4-1}
\usepackage{graphicx}
\usepackage{bm}
\usepackage[colorlinks=true,allcolors=blue]{hyperref}
\usepackage{dcolumn}

\usepackage{xcolor}
\definecolor{lightgray}{gray}{0.95}

\begin{document}
\title{The Critical Properties of the Ising Model in Hyperbolic Space}
	\author{Nikolas P. Breuckmann}
	\affiliation{Department of Physics \& Astronomy, University College London, WC1E 6BT London, United Kingdom}
	\email{n.breuckmann@ucl.ac.uk}
	\author{Benedikt Placke}
	\affiliation{JARA-Institute for Quantum Information, RWTH Aachen University, 52056 Aachen, Germany}
	\affiliation{Max-Planck-Institut f\"ur Physik komplexer Systeme, N\"othnitzer Str. 38, 01187 Dresden, Germany}\email{placke@pks.mpg.de}
	\author{Ananda Roy}
	\affiliation{Department of Physics, Technische Universit\"at M\"unchen, 85748 Garching, Germany}\email{ananda.roy@tum.de}
\begin{abstract}
The Ising model exhibits qualitatively different properties in hyperbolic space in comparison to its flat space counterpart. Due to the negative curvature, a finite fraction of the total number of spins reside at the boundary of a volume in hyperbolic space. As a result, boundary conditions play an important role even when taking the thermodynamic limit. We investigate the bulk thermodynamic properties of the Ising model in two and three dimensional hyperbolic spaces using Monte Carlo and high and low-temperature series expansion techniques. To extract the true bulk properties of the model in the Monte Carlo computations, we consider the Ising model in hyperbolic space with periodic boundary conditions. We compute the critical exponents and critical temperatures for different tilings of the hyperbolic plane and show that the results are of mean-field nature. We compare our results to the `asymptotic' limit of tilings of the hyperbolic plane: the Bethe lattice, explaining the relationship between the critical properties of the Ising model on Bethe and hyperbolic lattices. Finally, we analyze the Ising model on three dimensional hyperbolic space using Monte Carlo and high-temperature series expansion. In contrast to recent field theory calculations, which predict a non-mean-field fixed point for the ferromagnetic-paramagnetic phase-transition of the Ising model on three-dimensional hyperbolic space, our computations reveal a mean-field behavior.
\end{abstract}
\maketitle

\section{Introduction}
The critical properties of a statistical mechanics model on curved space can be drastically different from its flat-space counterpart~\cite{Callan1990}. In particular, statistical mechanics on negatively curved hyperbolic space has attracted much attention. First, they are relevant for quantum field theories in curved space-time~\cite{Wald1994} and serve as means to separate the roles of geometric curvature and topological defects~\cite{Callan1990}. Second, they arise in condensed matter settings as toy models for amorphous solids or exotic liquid crystalline structures~\cite{Kleman1982, Rubinstein1983, Nelson2002}. Third, hyperbolic spaces arise in quantum information theory as natural candidates for encoding quantum information with toric codes. This is because a toric code on hyperbolic space encodes a larger number of logical qubits than a flat-space counterpart for the same number of physical qubits~\cite{Freedman2002, Breuckmann2017}.

An essential consequence of negative curvature in a hyperbolic space is the exponential growth of the volume with an associated linear dimension.  This leads to the expectation that the space is effectively infinite-dimensional and the critical properties of any statistical mechanical model will be effectively mean-field. At first sight, this might lead to the impression that the thermodynamical properties of the hyperbolic-space statistical model can be straightforwardly inferred from its flat-space counterpart. However, this is not true and the physics is qualitatively different due to the curvature of the embedding space. For instance, in contrast to the euclidean-space counterpart, the order-disorder phase-transition of the XY model on a 2D hyperbolic plane is driven by the fluctuations of topological defects, while the infrared properties of the spin-wave fluctuations are well-behaved~\cite{Callan1990}. Critical statistical mechanical systems show exponential decay of correlations~\cite{Mnasri2015} and new-phases with broken translational invariance can arise, which are absent in the flat-space counterparts. Mathematical proofs for the existence of such phases exist for models describing percolation~\cite{Benjamini1999, Benjamini2001, Baek2009} and ferromagnetic Ising models~\cite{Wu1996, Wu2000}.

For the ferromagnetic Ising model on a hyperbolic plane, high and low-temperature expansion had obtained mean-field exponents for susceptibility and magnetization~\cite{Rietman1992}. Interestingly, the transition temperature, obtained from the series expansion calculations for a self-dual lattice  was different from that of the 2D square-lattice Ising model. Since the Kramers-Wannier duality~\cite{Krammers1941} also holds for self-dual hyperbolic lattices, assuming convergence of the series up to the critical point, this indicates the existence of a second phase-transition at a temperature related by the duality relation [see below, Eq. \eqref{eqn:KWduality}]. Indeed, the existence of this second phase-transition for a hyperbolic plane with free boundary condition was later proved~\cite{Wu1996, Wu2000,Jiang2018}. This phase-transition separates the low temperature, purely ferromagnetic phase from an intermediate phase. In this intermediate phase, the system has broken translational invariance, where there are infinitely many, infinitely large clusters of magnetized spins. These clusters survive in the thermodynamic limit purely due to the negative curvature of the embedding space and cannot arise in ordinary euclidean space~\cite{Aizenman1980}. Further increase of temperature eventually causes the system to transition from the intermediate phase to the high temperature, disordered phase. It is this latter phase-transition that was found using series expansion methods. Additional evidence of mean-field behavior of magnetization was shown with corner-transfer-matrix renormalization group calculations~\cite{Krcmar2008}. Monte Carlo calculations of bulk thermodynamics (throughout the paper, by bulk thermodynamics, we mean thermodynamics of a model on a vertex-transitive graph) have so far only been done on open boundary lattices. To eliminate the effects of the large fraction of total spins being on the boundary arising due to the negative curvature, outer layers of these open lattices are removed, focussing the analysis to the central region~\cite{Shima2005}. While the boundary effects can be interesting on their own~\cite{d_Auriac2001}, they prevent the analysis of bulk behavior of the hyperbolic plane Ising model.

\begin{figure}
\includegraphics[width = 0.30\textwidth]{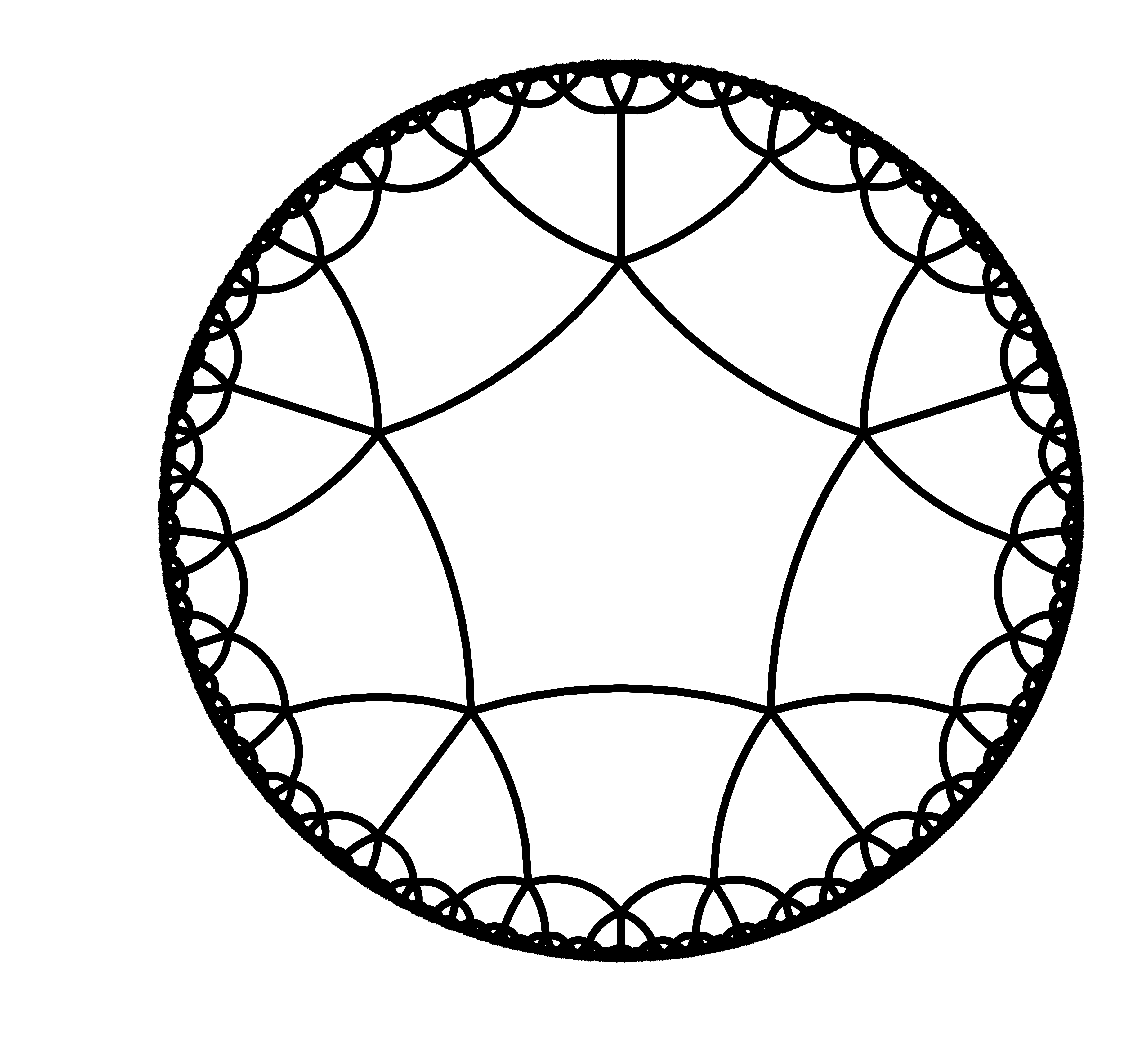}
\caption{\label{fig_ising} The self-dual $\{5,5\}$ tiling of the hyperbolic plane in the Poincar\'e disc model. Ising degrees of freedom reside on the nodes and any two spins connected by an edge interact. }
\end{figure}

In the first part of this work, we investigate the bulk critical properties of the Ising model on a hyperbolic plane. To remove boundary effects in the Monte Carlo simulations, we perform these simulations on a hyperbolic plane with periodic boundary conditions~\cite{Sausset2007}. We concentrate on the self-dual $\{5,5\}$ lattice, which is a tiling of the hyperbolic plane with regular pentagons (see Fig.~\ref{fig_ising}). We concentrate on this lattice since it captures all the qualitatively different physics of curved space. We compute the critical exponents and critical temperature as the system transitions to the ferromagnetic phase. A hyperbolic plane with periodic boundary conditions is a manifold with genus $\gg 1$. As a result, the finite size scaling is nontrivial and provides additional insight into the fundamental differences between statistical mechanical models defined on spaces with zero and negative curvature. We find that the universal critical exponents for the different thermodynamic quantities are close to the mean-field predictions. To lend additional credence to the above findings, we perform high and low temperature series expansion of various thermodynamic quantities. Performing this analysis, we find results that are close to our Monte Carlo predictions and confirm the mean-field nature of the phase-transition. Our results for the $\{5,5\}$ lattice are close to the results for the critical temperature recently obtained by Mone Carlo simulations of the same model~\cite{Jiang2018}.

Subsequently, we also compute the critical temperature and the critical exponents using Monte Carlo and series expansion for lattices with different  curvatures and compare our results for the different hyperbolic lattices to the Bethe lattice. We explain the relation between ferromagnetic-paramagnetic phase-transitions of the Ising model on hyperbolic lattices and those on the Bethe lattice, the latter being the limit where the number of edges of the polygons tiling the hyperbolic lattice is taken to infinity. Although not the point of the paper, we mention that we provide the high temperature series for several lattices not analyzed before, going up to ${\cal O} (v^{24})$, where $v = \tanh(\beta J)$. Here, $\beta$ is inverse temperature and~$J$ is the ferromagnetic coupling.

The last part of this work is devoted to the analysis of the Ising model in 3D hyperbolic space. Recent field theory calculations~\cite{Doyon2004, Benedetti2015} using 1/N expansion predict deviations from mean-field exponents for the Ising model in 3D hyperbolic space~\cite{Mnasri2015}. This result is remarkable since the basic intuition for critical properties of a statistical mechanical model in hyperbolic space being mean-field  --  that the hyperbolic space acts like an infinitely connected lattice -- is even more valid in 3D compared to 2D since the number of neighbors of each spin is higher. Previous calculations using series expansion or corner-transfer-matrix renormalization group have all been done for 2D hyperbolic Ising models. In this part of the work, we calculate the bulk critical properties of the model in 3D hyperbolic space with Monte Carlo and high-temperature series expansion technique. Our results, in contrast to the field theory calculations, reveal a mean-field nature of the transition.
We perform the Monte Carlo analysis of the Ising model on the $\{5,3,5\}$ lattice with periodic boundary condition, the latter being a tiling of 3D space with pentagonal dodecahedrons. Then, we compute the susceptibility exponents for the $\{5,3,4\}$ and $\{5,3,5\}$ lattices using high-temperature series expansion. We also consider a 3D analogue of the Bethe lattice which has Schl\"afli symbol $\{5,3,6\}$.

The article is organized as follows. In Sec.~\ref{sec:model}, we analyze the Ising model on a 2D hyperbolic plane. After a brief summary of the well-known properties of the model, in Sec.~\ref{sec:monte_carlo}, we perform Monte Carlo analysis to compute the universal critical exponents and infer the phase-transition point for the $\{5,5\}$ lattice. Then, we compute with high and low-temperature series expansions the susceptibility and the magnetization in Sec.~\ref{sec:series_expansion}. We find the results of the series-expansion computations to be close to those obtained by Monte Carlo. In Sec. \ref{sec:variation_with_curvature}, we describe the variation of the critical properties with curvature for self-dual lattices and explain the relationship between the critical properties of the Ising model in hyperbolic lattices and in the Bethe lattice. In Sec.~\ref{3Dmc}, we analyze the properties of the Ising model in 3D hyperbolic space with $\{5,3,5\}$ tiling using Monte Carlo. The high-temperature expansion calculations for the $\{5,3,k\}$ lattices, $k=4,5,6$ are done in Sec.~\ref{3dtemp}. Finally, in Sec.~\ref{sec:conclusion}, we provide a concluding summary and outlook. In Appendix~\ref{hyperbolic_tiling}, we summarize the relevant properties of hyperbolic planes and describe how to tessellate them in the presence of periodic boundary condition. Appendix~\ref{ap:series_derivation} provides additional details on the high-temperature series expansion. Appendix~\ref{sec:kramers_wannier} provides a general derivation of the Kramers-Wannier duality.

\section{The Ising Model on a 2D Hyperbolic Plane}
\label{sec:model}
Consider a regular tiling of the hyperbolic plane, denoted by $\{r,s\}$. Here, $r$ denotes the number of sides of the polygon and $s$ the coordination number of each site (more details in Appendix~\ref{hyperbolic_tiling}).
Note that for $r=s$ the lattice is self-dual; the case $r=s=5$ is shown in Fig.~\ref{fig_ising}. The Hamiltonian for the ferromagnetic Ising model on the hyperbolic plane is defined by
\begin{align}
\label{ham_ising}
H = -J \sum_{\langle i, j\rangle} \sigma_i\sigma_j,
\end{align}
where the spins ($\sigma_i = \pm1$) reside on the vertices of the lattice and any two spins connected by an edge interact with each other with a coupling $J>0$.

For either free, fixed or periodic boundary conditions, for temperatures $T\gg J$, the system is in a disordered, paramagnetic phase. Upon lowering the temperature, the system undergoes a phase-transition at a critical temperature, denoted by $T_c$, to a magnetically ordered phase. For free boundary conditions, in contrast to the euclidean space Ising model, this phase has broken translational invariance and is characterized by {\it infinitely many} clusters of magnetized spins. The survival of these clusters in the thermodynamic limit is due to the negative curvature of the hyperbolic plane, which allows an infinite number of these clusters to be squeezed within the plane. For free boundary conditions, upon further lowering of the temperature, at a temperature $\bar{T}_c$, the system undergoes another phase-transition, where a {\it single} cluster of magnetized spins covers the entire hyperbolic plane and translational invariance is restored. The two temperatures, $T_c, \bar{T}_c$, are related to one another by the Kramers-Wanner duality relation:
\begin{align}\label{eqn:KWduality}
\sinh(2J/T_c)\sinh(2J/\bar{T}_c) = 1,
\end{align}
(see Appendix \ref{sec:kramers_wannier} for a proof of the duality for all self-dual lattices).
This qualitative difference between Ising models in absence of external magnetic field for planes with and without curvature can be mathematically formulated in a unified framework using the cluster model, which describes both percolation and Ising models as special cases. 
The proof for free boundary condition is formulated in terms of the (non-)uniqueness of the Gibbs states~\cite{Wu1996, Wu2000} and the intuitive explanation is given below. Consider the Gibbs states $\nu_\pm, \nu_f$ for the Ising model with fixed, free boundary conditions. Here, $+(-)$ refers to the fixed boundary condition case when the boundary spins are fixed to $+1(-1)$ states. First, consider planes without curvature. In the high-temperature phase ($T\gg J$), the Gibbs state is unique: $\nu_+ = \nu_- = \nu_f$. The average magnetization ($m$) is zero in this phase and the correlation function $\langle\sigma_i \sigma_j\rangle_{f,\pm}\rightarrow0$ for $|i-j|\rightarrow \infty$, the decay to zero being exponential due the presence of a gapped spectrum. On the other hand, in the low-temperature phase ($T\ll J$), the Gibbs-state is non-unique $\nu_+ \neq \nu_-$; However, these two states remain extremal, which means that the Gibbs state with free boundary condition is a symmetric mixture of the two:~$\nu_f = (\nu_+ + \nu_-)/2$. In this phase, the absolute-value of the magnetization is nonzero and $\langle\sigma_i \sigma_j\rangle_{f,\pm}\geq m^2$. The two phases are separated by a second-order phase-transition. Now, consider the Ising model on the hyperbolic plane. The high-temperature phase ($T\gg J$) again has unique Gibbs states. The average internal magnetization is zero and correlations are exponentially damped. Upon lowering the temperature, at $T_c$, the system magnetizes. However, the Gibbs states, $\nu_\pm$, are no-longer extremal, which leads to~$\nu_f \neq (\nu_+ + \nu_-)/2$. As a result, there exists a finite fraction of uncorrelated spins, which survive in the thermodynamic limit:~$\langle\sigma_i \sigma_j\rangle_{f}\rightarrow0$, despite there being a finite overall magnetization. Finally, upon further lowering of temperature, at $\bar{T}_c$, the system is magnetized as a whole and the extremal nature of the Gibbs states~$\nu_\pm$ is restored. This leads to~$\nu_f = (\nu_+ + \nu_-)/2$ and~$\langle\sigma_i \sigma_j\rangle_{f,\pm}\geq m^2$. This low-temperature phase is similar to the low-temperature phase of the euclidean space Ising model. 

We emphasize that the phase-diagram of the model depends on the different choice of boundary conditions. The existence of the intermediate phase has been proven for free boundary conditions. We do not know if a similar intermediate phase exists for the case of periodic boundary conditions; see Sec. \ref{intermphase} for further discussion. Note that, despite the presence of two phase-transitions, the magnetic susceptibility diverges and the magnetization vanishes only at the higher phase-transition temperature Tc. These critical properties of these thermodynamic quantities at this higher temperature phase-transition are analyzed using Monte Carlo and series expansion methods below.

\subsection{Monte Carlo analysis of the critical exponents and temperature}
\label{sec:monte_carlo}
\begin{figure}
\centering
\includegraphics[width = 0.5\textwidth]{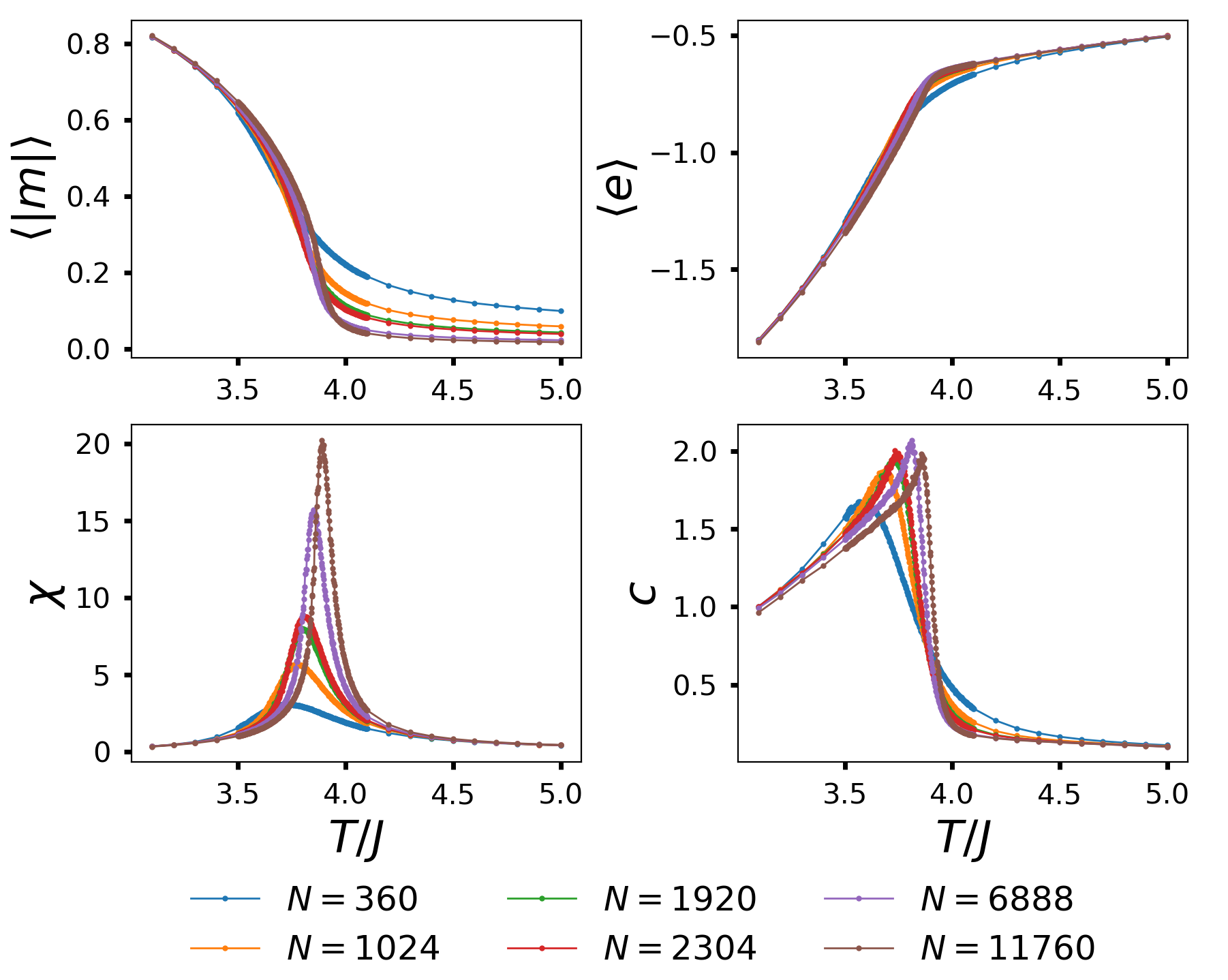}
\caption{Results of the Monte Carlo simulations for the hyperbolic space with $\{5,5\}$ tiling. The absolute magnetization per spin ($|m|$), the energy per spin ($\langle e\rangle$), average absolute susceptibility per spin ($\chi$) and the specific heat per spin ($c$) are plotted  in the top left, top right, bottom left and bottom right panels respectively. We only show a zoom of the different quantities near the phase-transition region; but simulations were done from deep in the ferromagnetic regime to deep in the disordered region.}
\label{fig:mc_1}
\end{figure}

\begin{figure}
\centering
\includegraphics[width = 0.5\textwidth]{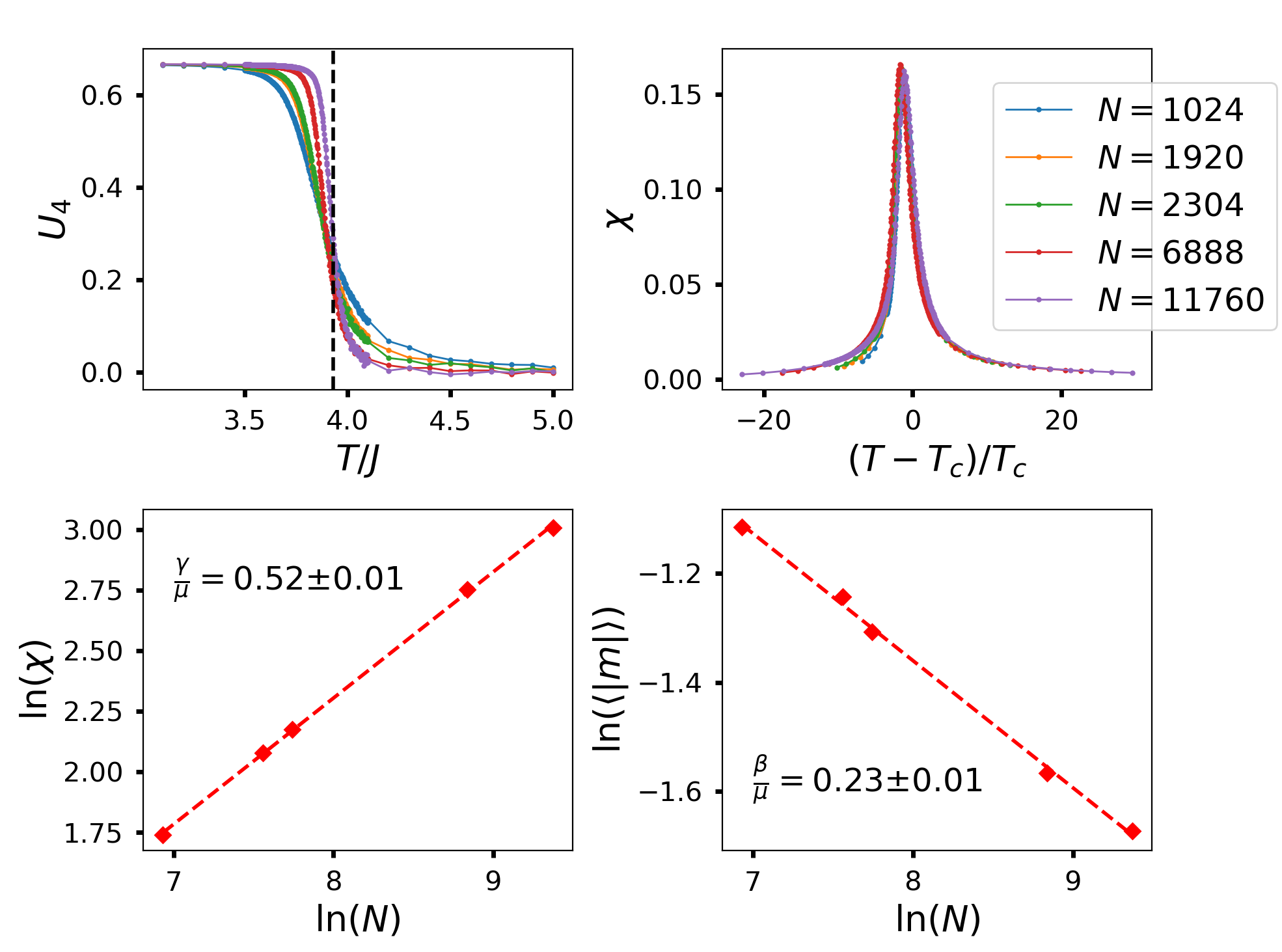}
\caption{\label{fig_mc_2} Results of the Monte Carlo simulations: (top left) the fourth Binder cumulant, (top right) data collapse for average absolute susceptibility and the linear fits are system size scaling for average absolute magnetization (bottom left) and average absolute susceptibility per spin (bottom right). The vertical line in the top left panel $T/J = 3.93$ (see main text for error estimate) indicates the location of phase-transition temperature obtained from the data collapse. The linear fits in the bottom panels denote are obtained from the finite size analysis with respect to the number of spins. The slope for the linear fit for average absolute susceptibility (magnetization) per spin provides an estimate of the ratio $\gamma/\mu$ ($\beta/\mu$), where $\gamma$ ($\beta$) denote the exponents for susceptibility (magnetization) and $\mu$ is the scaling of the coherence number. In the bottom panels, the error indicated is only the fit error, the actual value and the error in the exponent is provided in the main text. }
\end{figure}

In this subsection, we perform Monte Carlo analysis of the Ising model on the $\{5,5\}$ lattice using the Metropolis algorithm.
In order to perform finite-size scaling we perform the numerical analysis for perfiodic lattices of different size.
We compute the average  magnetization per spin $m = M/N$ and the energy per spin $e = E/N$, where~$M$ is the total magnetization,~$E$ the total energy and~$N$ the number of spins in the lattice.
Then, we compute the average absolute susceptibility per spin $\chi$ and the specific heat per spin $c$, defined below~\cite{Newman1999}
\begin{align}
\chi &= \frac{\beta}{N}(\langle M^2\rangle - \langle |M|\rangle^2), \\
\ c &= \frac{\beta^2}{N}(\langle E^2\rangle - \langle E\rangle^2),
\end{align}
where $\beta=1/T$ is the inverse temperature and $k_B$ is the Boltzmann constant. The phase-transition point is inferred form the fourth Binder cumulant~\cite{Landau2009}, given by
\begin{align}
U_4 = 1-\frac{\langle m^4\rangle}{3\langle m^2\rangle^2}.
\end{align}
Each Monte Carlo simulation was equilibriated with $10^5$ sweeps of the lattice, where one sweep corresponds to the number of Metropolis updates equal to the number of spins in the lattice. The measurements were performed over $10^6$ sweeps. The simulations were done for lattices with the following number of spins: $N = 360$, $1024$, $1920$, $2304$, $6888$ and $11760$. Fig. \ref{fig:mc_1} shows the absolute magnetization per spin, the average energy per spin, the average absolute susceptibility per spin and the specific heat per spin. The error analysis was performed using the bootstrap method~\cite{Newman1999}. The computed errors in the measured quantities are too small to show in the plots. The maximum of the absolute susceptibility occurs at the same temperature at which the magnetization goes to zero, in this case indicating a transition from a magnetically ordered phase to a disordered, high temperature phase. A crude estimate of the location of this phase-transition point can be obtained with help of the fourth Binder cumulant $U_4$, plotted in the top left panel of Fig. \ref{fig_mc_2}.

Accurate estimates of the phase-transition temperature and the critical exponents require finite system size analysis. In simulations on a finite plane with zero curvature, the divergence of the correlation length is cut off by the linear dimension of the system. From standard homogeneity arguments~\cite{Newman1999, Landau2009}, one can then obtain the ratio of the critical exponents: {\it e.g.}, from a linear fit of $\ln\chi$ versus $\ln L$, one gets the ratio $\gamma/\nu$, where $\gamma$ and $\nu$ are the critical exponents for the susceptibility and the correlation length and $L$ is the linear system size. Similarly, the the ratio $\beta/\nu$ is obtained from the linear fit of $\ln \langle |m|\rangle$ versus $\ln L$, where $\beta$ is the critical exponent for magnetization. The exact numerical values of the exponents and the phase-transition point are extracted by collapsing the data. This situation is different for a hyperbolic plane for following reason. In order to remove boundary effects, we perform the Monte Carlo simulations on a compactified hyperbolic plane, which is a manifold with genus larger than one. 
The size of the smallest non-contractible loop along different handles this manifold may vary.\footnote{Note that this is not in contradiction to the space being isotropic, as several non-contractible loops my run through the same vertex (or the same edge) and these loops may  have different lengths.} 
As a result, the choice of a suitable linear dimension is ambiguous.

To combat these difficulties of finite size scaling, we choose to perform finite size scaling with respect to the number of spins, which was initially proposed for an infinitely connected lattice~\cite{Botet1982} and has been used for hyperbolic space Ising models with open boundaries \cite{Shima2005}. We define a correlation number $N_c$, which plays the role of the correlation length in flat space, with an assumed scaling: $N_c\propto |T-T_c|^{-\mu}$, where $\mu$ is the associated critical exponent and $T_c$ is the critical temperature. The exponent $\mu$ is given by $\nu_{\rm{MF}}d_c$~\cite{Botet1982}, where $\nu_{MF}$ is the mean-field exponent of the divergence of correlation length for the Ising model ($=1/2$) and $d_c$ is the upper critical dimension ($=4$). Thus, $\mu = 2$. As explained in Ref.~\cite{Botet1982}, the correlation number $N_c \sim \xi^{d_c}$, where $\xi$ is the correlation length of the Euclidean space model. The use of an effective correlation number, which scales with the overall volume of the space, avoids the ambiguities associated with the linear dimension. The rest of the finite size scaling is similar to that of euclidean space, with the linear dimension replaced by the number of spins and the correlation length exponent replaced by $\mu$. Physically, due to the exponential growth of hyperbolic space volume with linear dimension, we expect the finite-size scaling analysis used for infinitely connected lattices to be applicable in our case. Since there is no proof that such an analysis is indeed valid, in Sec. \ref{sec:series_expansion}, we compute the critical exponents and critical temperature using high and low-temperature expansions. We do this since the series expansion computations, by design, do not suffer from finite size effects (but have different sources of error, see below). We find the results of these two independent computations to be close to each other, which lends additional credence to our Monte Carlo findings.

Performing this analysis, we find that the critical transition temperature occurs at $T_c / J = 3.93\pm0.03$. The critical exponents are $\beta = 0.46\pm 0.05$ and $\gamma = 1.03\pm 0.02$. The critical temperature, $T_c$, is close to the two possible candidates for critical temperature given in Ref. \cite{Jiang2018}. The exponents obtained are close to those for the Ising model within the mean-field approximation.  The divergence of the specific heat does not develop as the system size is increased and we expect the critical exponent $\alpha$ to be 0, as expected for a mean-field theory. We note our Monte Carlo results are not exactly the mean-field predictions, despite being close to them. We attribute the deviations to the finite-size scaling done by considering a divergence of a correlation number. Additional support for the mean-field behavior is obtained by high and low temperature series expansion computations, described below.

\subsection{High-temperature expansion analysis of the critical exponents and temperature}
\label{sec:series_expansion}
In this section, we perform high-temperature series expansion of the susceptibility on the infinite lattice.
We chose to concentrate on the susceptibility rather than the free energy density as we found the latter harder to analyze.
It turns out that it is favourable to perform the high-temperature expansion in the {\em inverse} susceptibility.
The reason for this is that it can be shown~\cite{singh87} that the only non-trivial contributions come from graphs which have the property that they stay connected if any of their vertices (and the edges attached to it) are being removed.
Such graphs are called {\em biconnected} graphs and since it is a more restricted class there are fewer of these graphs which simplifies the expansion and allows us to go to much higher orders.

The inverse susceptibility can be expanded in terms of these graphs as
\begin{align}\label{eqn:sus_series}
\tilde{\chi}^{-1} = 1\, + \, \sum_{g}\, c(g)\, W(g)
\end{align}
where the sum is over all graphs, $c(g)$ is the coefficient of~$N$ of the number of embeddings of the graph~$g$ into the lattice and~$W$ is a weight function. For more details and a derivation of Eq.~(\ref{eqn:sus_series}) see Appendix~\ref{ap:series_derivation}.

\begin{figure}
	\centering
	\includegraphics[width = 0.45\textwidth]{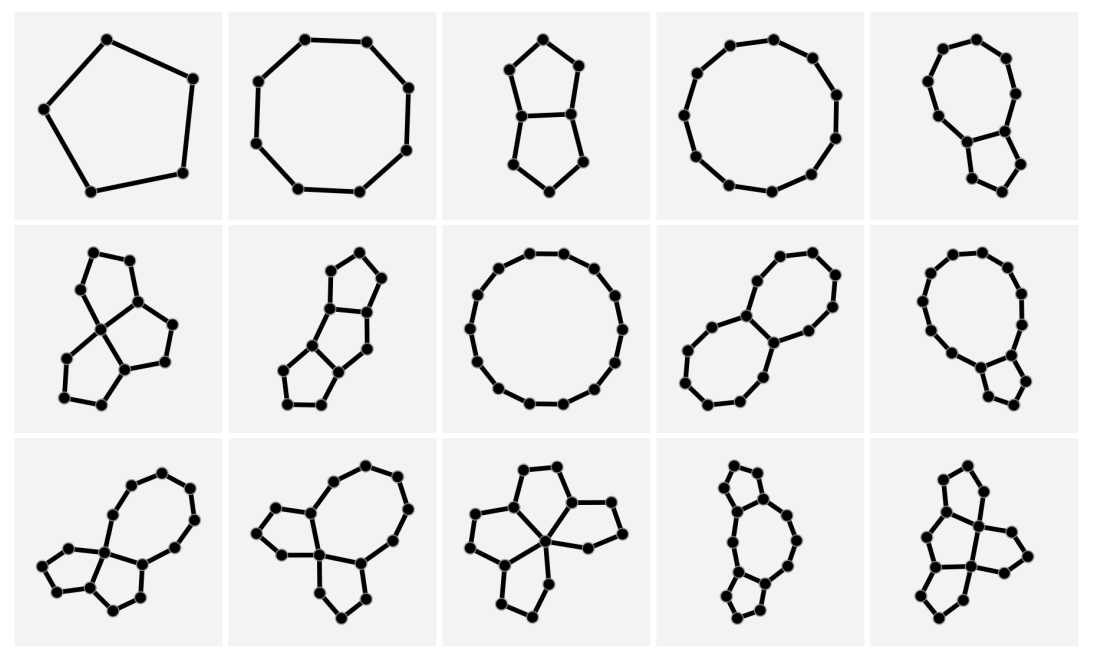}
	\caption{Some small biconnected subgraphs of the $\{5,5\}$-tiling.  Removing a vertex and all its incident edges will leave the graphs connected. Only biconnected graphs contribute to the series expansion.}\label{fig:55bicon}
\end{figure}

\subsubsection{Analysis of the Series}
The coefficients of the susceptibility on the $\{5,5\}$-lattices are given in Table ~\ref{tab:susseries}.
Our results for the high-temperature series expansion agree with \cite{Rietman1992} who obtained the series for~$\{5,5\}$ up to order~10.

We analyze the series $\tilde{\chi}(v)$ using \emph{first-order homogeneous integrated differential approximants (FO-IDAs)}.
One reason to choose FO-IDAs over simpler methods is that they are known to be less biased towards the lower-order coefficients of the expansion~\cite{singh2}.
This is important here, as the non-trivial contributions come from graphs with at least~$r$ edges.

The analysis using FO-IDAs proceeds as follows:
We assume that the series $\tilde{\chi}$ is the solution of a first-order differential equation of the form
\begin{equation}\label{eqn:IDAdef}
Q_L(v) \frac{d \tilde{\chi}(v)}{d v} + R_M(v)\, \tilde{\chi}(v) + S_T(v) = 0
\end{equation}
where $Q_L$, $R_M$ and $S_T$ are polynomials of degree $L$, $M$, $T$, respectively.
By equating the series order-by-order with the coefficients of Eq.~\ref{eqn:IDAdef} we obtain a linear system of equations in the coefficients of the polynomials $Q_L$, $R_M$ and $S_T$.
It can be shown that for any root $v_c$ of the polynomial~$Q_L$, a solution of Eq.~\ref{eqn:IDAdef} has an algebraic singularity of the form $(v-v_c)^{-\gamma}$ \cite{oitmaa2006}.
The exponent of the singularity is given by
\begin{align}\label{eqn:crit_exp}
\gamma = \frac{R_{M}(v_c)}{Q'_L(v_c)} .
\end{align}
Generally, the results for $v_c$ and $\gamma$ will depend on the choice of degrees $L$, $M$ and $T$.
If we have the series up to order $N$ then we can choose all possible values satisfying $L+M+T \leq N-2$.
Following~\cite{singh2} we exclude approximants if
\begin{itemize}
	\item a root of $R_M$ is close to $v_c$, giving rise to a small estimate of $\gamma$
	\item a complex root of $Q_L$ with small absolute value smaller than $v_c$ is close to the real axis
\end{itemize}

We observe that the convergence of the series is very good, since the approximants for different choices of $L$, $M$ and $T$ are all close.
The mean critical value of $v$ obtained by averaging over 45 approximants is $v_c = 0.25200759 \pm 0.00000006$.
This is in agreement with the result obtained by Monte Carlo [$\tanh(1/3.93) \approx 0.249$].
The critical exponent $\gamma$ is found via Eq. \ref{eqn:crit_exp} and averaged over all approximants.  The result of $\gamma = 1.000001 \pm 0.000005$ is again close to the Monte Carlo result and the mean field value $\gamma = 1$.

\subsubsection{Comparison to low-temperature series}

\begin{figure}
	\includegraphics[width = 0.45\textwidth]{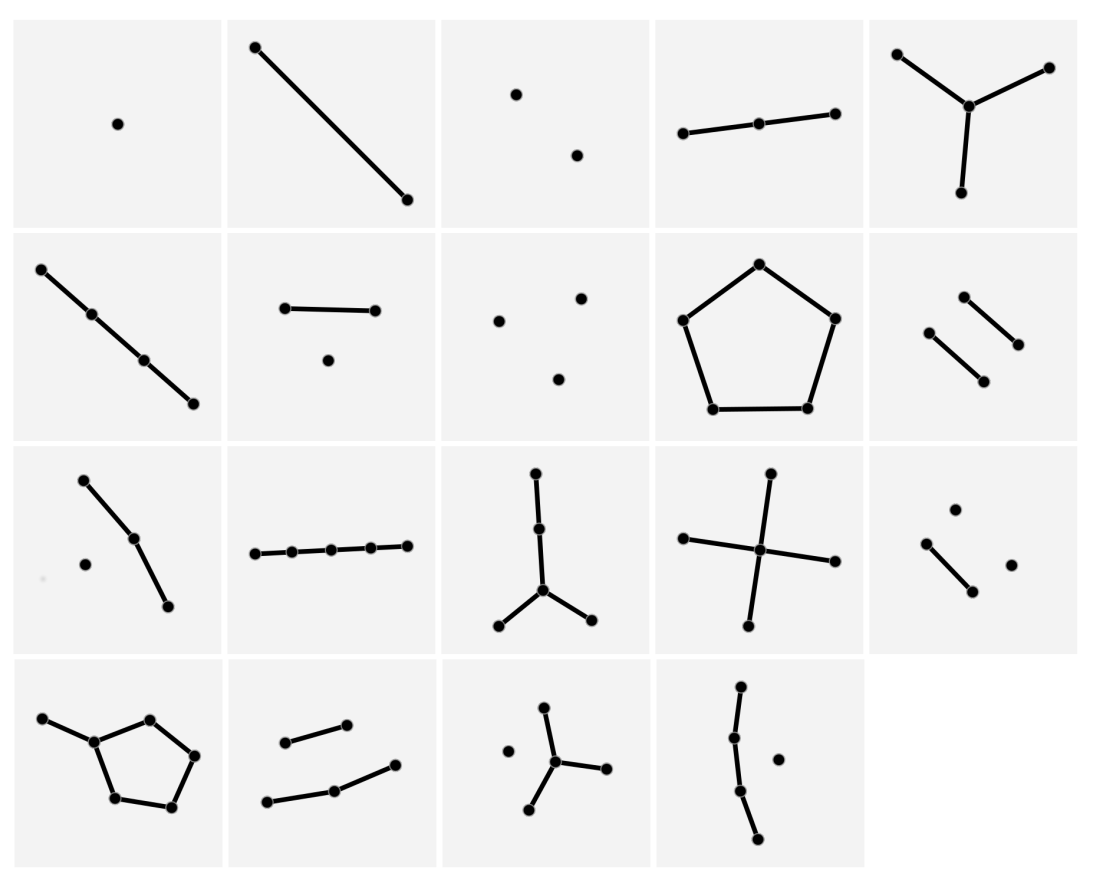}
\caption{All the graphs contributing up to order $u^{19}$ of the magnetization of the model on the $\{5,5\}$-tiling.}\label{fig:55all}
\end{figure}

To verify that thermodynamic properties only diverge at the higher critical temperature $T_c$, we perform also a low temperature series of the magnetization of the model on the $\{5,5\}$ lattice. In the case of the low temperature expansion, we use an elementary expansion technique. We write the free energy in the presence of a field as the sum of configurations of flipped spins with respect to the ground state, to obtain an expansion in the temperature $u = \exp(-2\beta J)$ and the field $\mu = \exp(-2 \beta h)$
\begin{equation}\label{eqn:low_temp_Z}
    \frac{1}{N} \ln Z = \frac{\ln(2)}{N} + \frac{q}{2} \beta J + \beta h + \sum_{\{g\}} c(g) u^{q n_v-2n_l} \mu^{n_v},
\end{equation}
where $q$ is the coordination number, $h$ is the magnetic field and $g$ is any graph on the lattice (including articulated and disconnected graphs). $c(g)$ here is the part of the number of embeddings of the graph $g$ into the lattice that is proportional to $N$, and $n_v$ and $n_l$ are the number of vertices and edges of the graph respectively.
We note that setting $h=0$ ($\mu=1$) we obtain the exact same coefficients as for the high-temperature expansion by exchanging $u \leftrightarrow v$, as predicted by the Kramers-Wannier duality (cf. Appendix~\ref{sec:kramers_wannier}).
From Eq.~\ref{eqn:low_temp_Z} the magnetization is obtained via
\begin{equation}
    M = -2 \mu \frac{\partial}{\partial \mu} \frac{1}{N} \ln Z.
\end{equation}

All graphs contributing to the magnetization on the $\{5,5\}$ lattice up to order $u^{19}$ are shown in Fig. \ref{fig:55all}. Summing up their contributions yields:
\begin{align}
    M = 1 &- 2u^5 \mu - 10 u^8 \mu^2 + 12 u^{10} \mu^2 - 60 u^{11} \mu^3 \nonumber\\
    &+ 150 u^{13} \mu^3 - 400 u^{14} \mu^4 - (92\mu^3+10\mu^5)u^{15}\nonumber\\
    &+ 1530\mu^4 u^{16} - 2800\mu^5u^{17} - (1920\mu^4+180\mu^6)u^{18}\nonumber\\
    &+14600\mu^5u^{19}+ \mathcal{O}(u^{20})
\end{align}
The magnetization is expected to vanish at the critical temperature with a power law
\begin{equation}
    M \sim (T_c - T)^\beta.
\end{equation}
We can thus analyze the inverse magnetization using FO-IDAs as before. This yields a critical point
\begin{equation}
    u_c = 0.608 \pm 0.014
\end{equation}
which corresponds to a upper critical temperature of $T_c = 4.01\pm0.19$. This is in agreement with both, the result from the high-temperature series and the Monte Carlo simulation. Also, in agreement with the Monte Carlo simulation, we see that the magnetization does not vanish at the lower critical temperature $\bar{T}_c$. Finally, we note that our results are commensurate with those obtained in Ref. \onlinecite{Rietman1992}.

\section{Variation of critical properties with curvature}\label{sec:variation_with_curvature}
In this section, we perform Monte Carlo and series expansion analysis of Ising models on hyperbolic planes with different (but uniform) curvatures.
Before we do so we explain how the lattice type $\{r,s\}$ is connected to the magnitude of curvature of the underlying space.

\subsection{Curvature and lattices}
One way to quantify curvature is by considering a circle centered around a point $p$ with radius $r$.
Let~$C_p(r)$ be the circumference of this circle.
The curvature around the point $p$ is the difference between the circumference~$C_p(r)$ and the circumference of a circle with the same radius in the euclidean plane
\begin{align}
	\kappa_p = \lim_{r \rightarrow 0^+} 3\; \frac{2\pi r - C_p(r)}{\pi r^3}.
\end{align}
Intuitively, the curvature at $p$ is negative if there is an excess of space in a local neighborhood around it.
This excess of space allows for a large variety of lattices in hyperbolic space.
\begin{figure}
	\centering
	\begin{minipage}{.6\columnwidth}
		\centering
		\includegraphics[width=0.45\columnwidth]{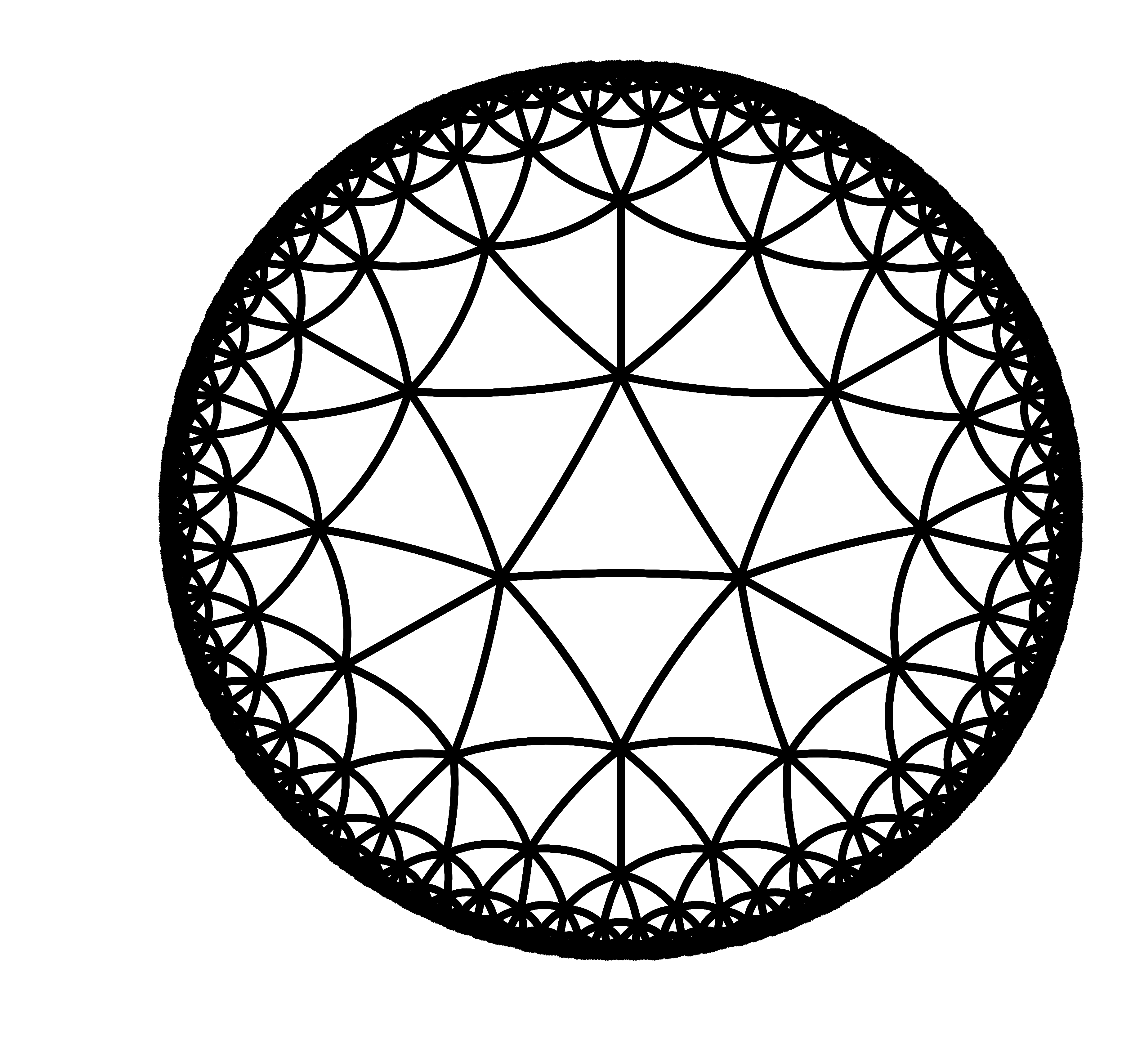}
		\includegraphics[width=0.45\columnwidth]{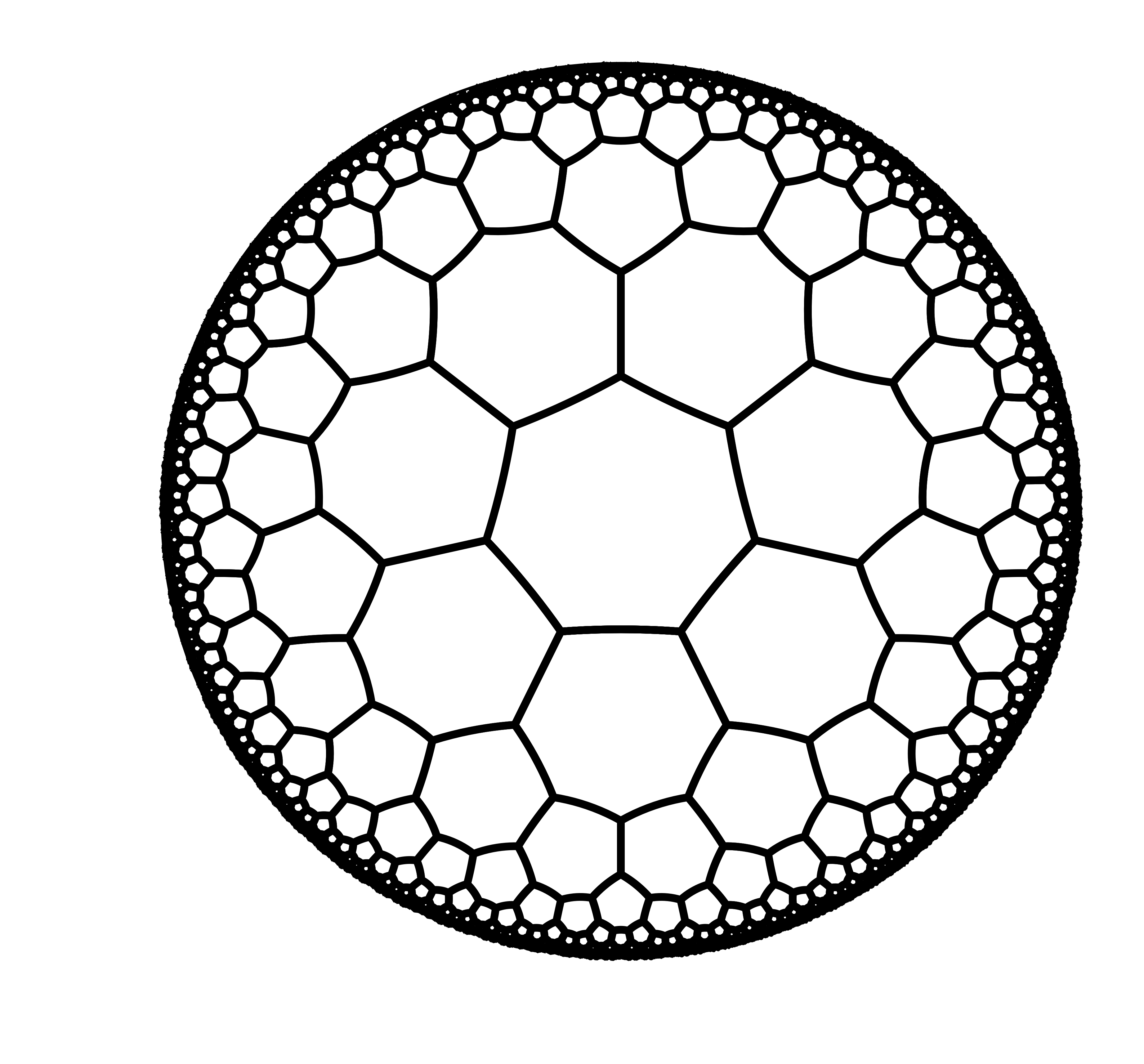} \\
		\includegraphics[width=0.45\columnwidth]{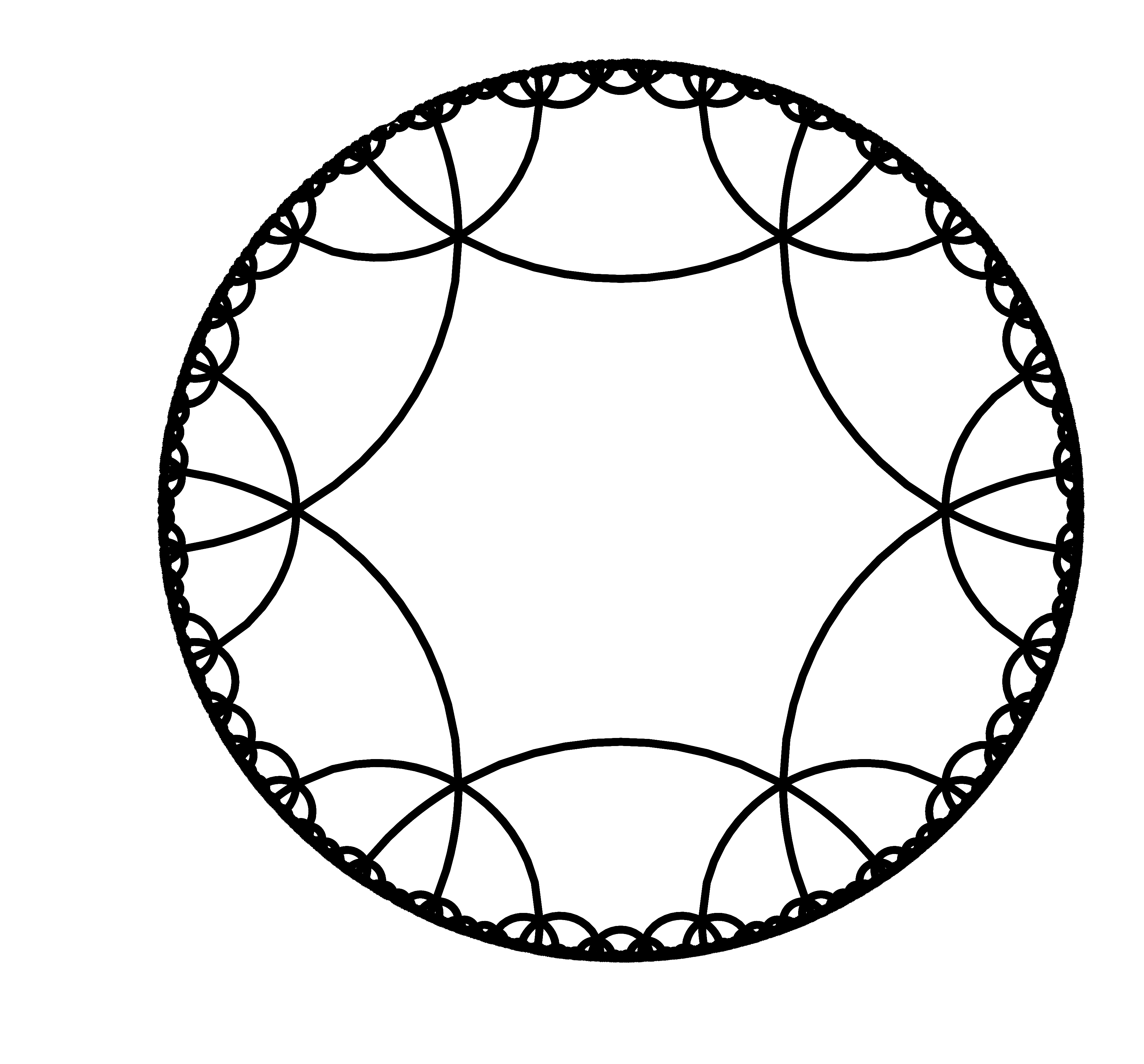}
		\includegraphics[width=0.45\columnwidth]{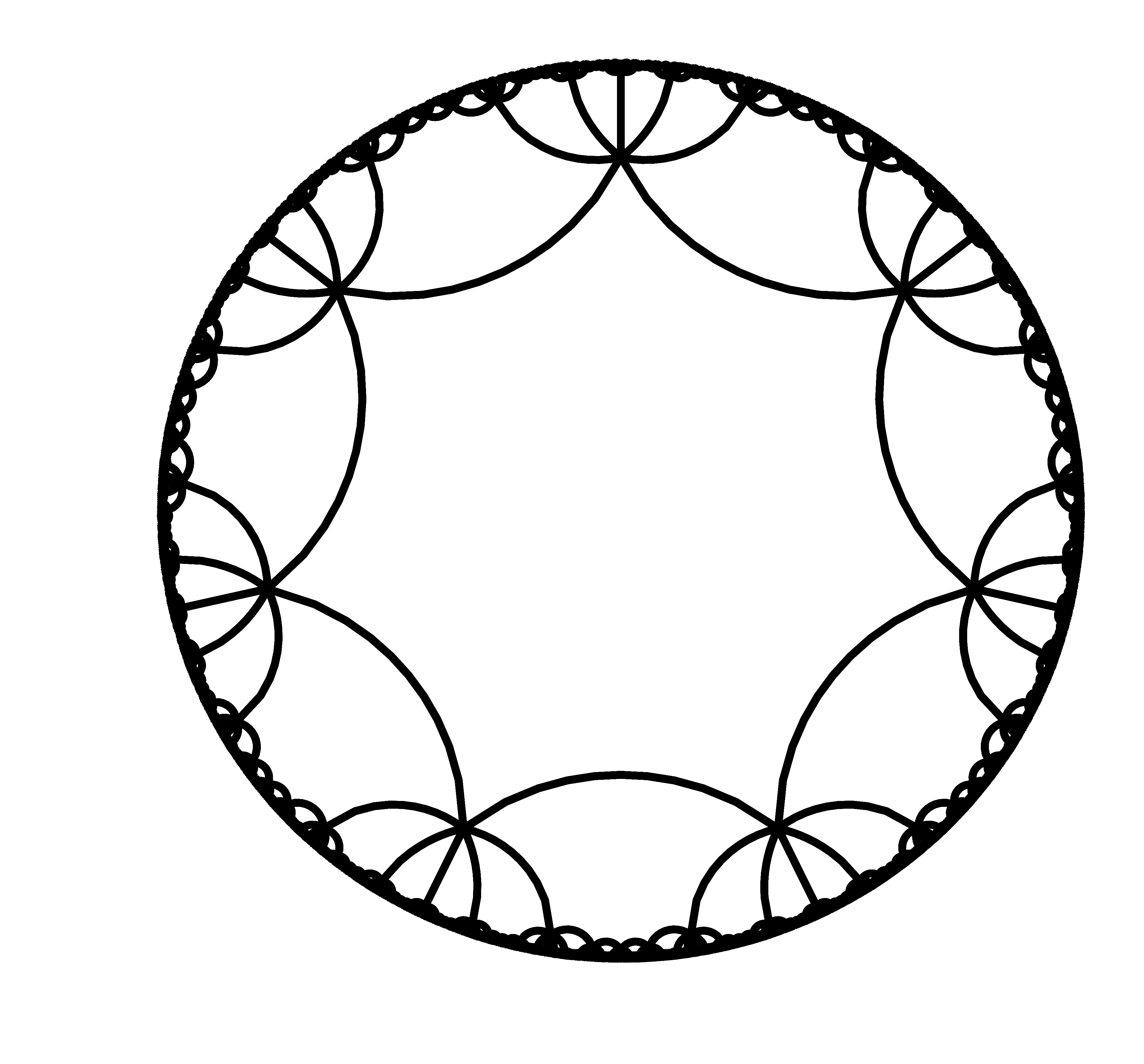}
	\end{minipage}%
	\begin{minipage}{.4\columnwidth}
		\centering
		\includegraphics[width=0.7\columnwidth]{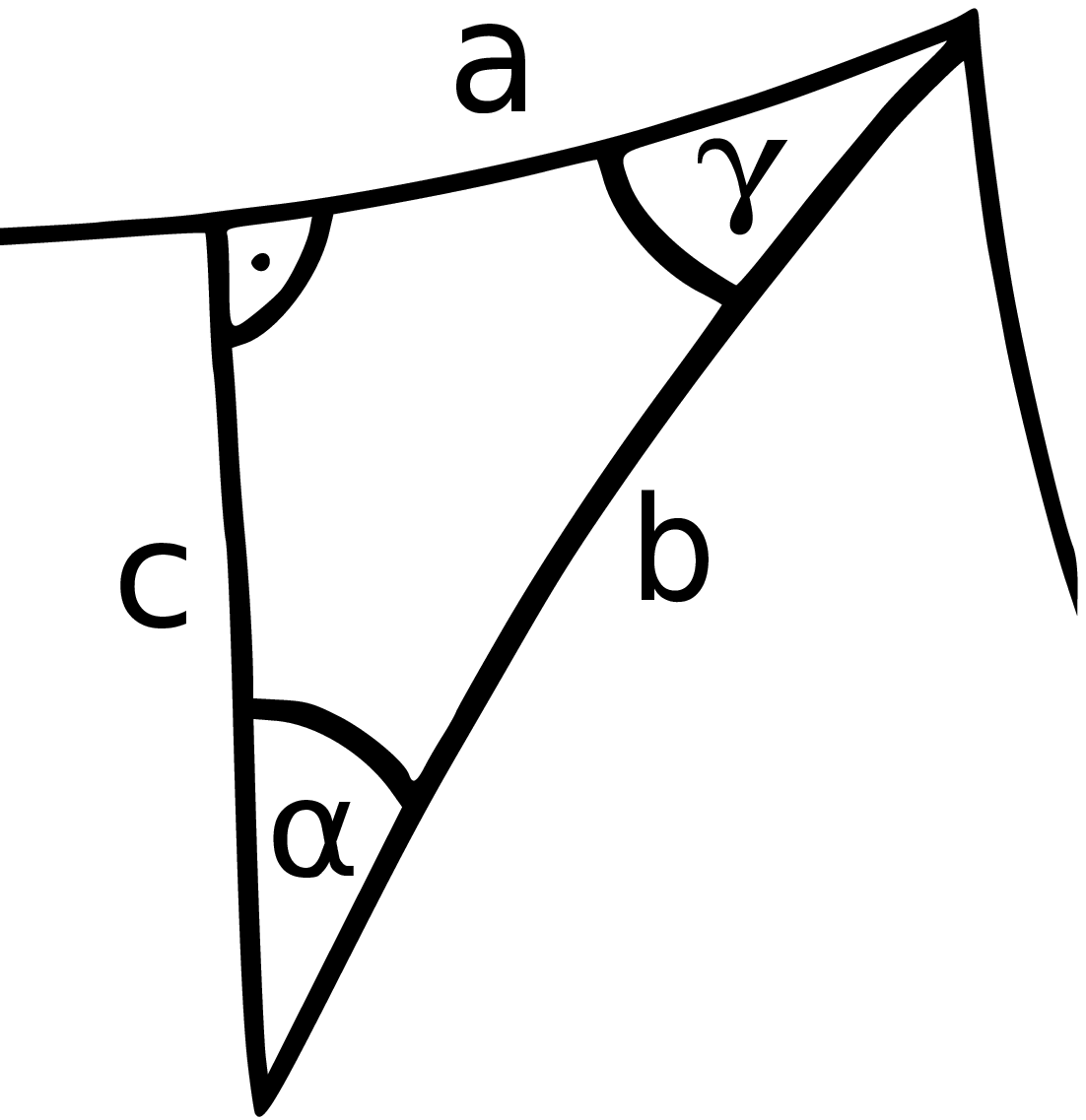}
	\end{minipage}

	\caption{Left: More tilings of the hyperbolic plane. Their Schl\"afli symbols from top to bottom and left to right: $\{3,7\}$, $\{7,3\}$, $\{6,6\}$, $\{7,7\}$. The tilings in the upper row are dual to one another, while the tilings in the bottom row are self-dual. Right: Angles of a triangle in the lattice.}
	\label{fig:hyptilings}
\end{figure}
In fact, curvature and the type of regular tiling are intimately connected.
This can be seen as follows.
If we normalize the edge-length to be $1$, the hyperbolic law of cosines states that for any triangle in the hyperbolic plane with internal angles $(\alpha, \beta, \gamma)$ and $a$ the side-length opposing $\alpha$ we have
\begin{align}\label{eqn:hypcosinelaw}
	\cos(\alpha) = -\cos(\beta)\, \cos(\gamma) + \sin(\beta)\, \sin(\gamma)\, \cosh(\kappa a) .
\end{align}
Triangulating a face of the tessellation by drawing lines from the center of the face to a vertex and the mid-point of an edge gives a triangle with angles $(\alpha,\beta,\gamma) = (\pi/r,\pi/2,\pi/s)$ and $a=1/2$ (see Fig.~\ref{fig:hyptilings}).
It now follows that
\begin{align}\label{eqn:schlafli_curvature}
	\kappa = -4\, \cosh^{-1}\left(\frac{\cos(\pi/r)}{\sin(\pi/s)}\right) .
\end{align}

\subsection{Monte Carlo and series expansion analysis}
Next, we perform Monte Carlo and high-temperature series expansion analysis of the Ising model for the following lattices: $\{r,r\}$, $r = 5,6,7,8$. As the coordination number, $r$, increases, the curvature of the plane also increases [cf. Eq.~\eqref{eqn:schlafli_curvature}]. This causes the phase-transition temperature to increase as well. This is because increase in the coordination number increases the effective strength of the interaction seen by each spin.

\begin{figure}
	\centering
	\includegraphics[width = 0.5\textwidth]{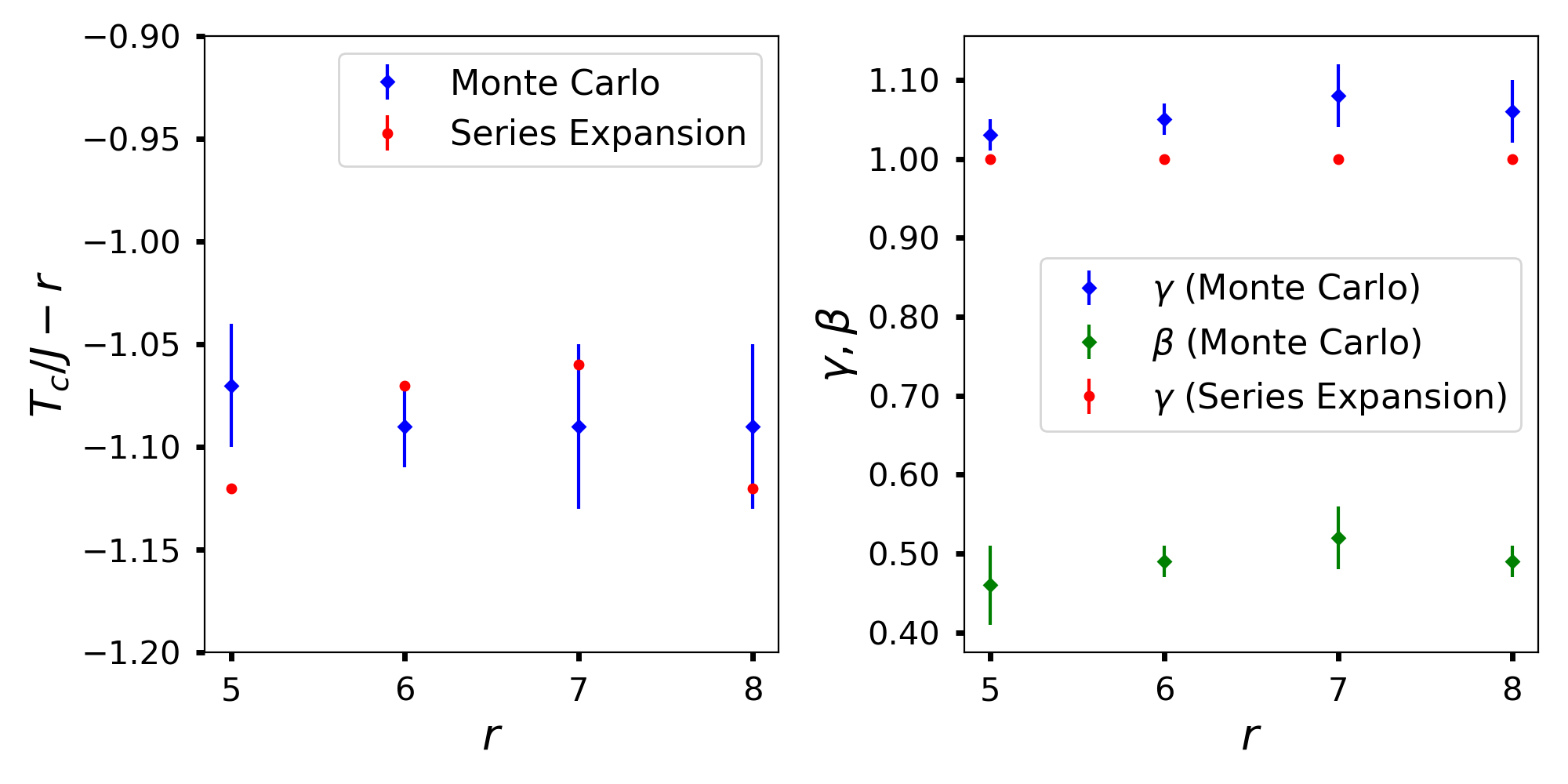}
	\caption{\label{fig:mc_variation} Results for self-dual $\{r,r\}$ lattices from Monte Carlo (filled blue diamonds) and high-temperature series expansion (filled orange circles) computations.  (Left panel) The obtained transition temperatures, $T_c$ (in units of $J$), with the coordination numbers, $r$, subtracted to ensure the visibility of the error bars.
	We note that the transition temperatures are increase by the same amount as the coordination number is increased by 1. This behavior is further indication that the nature of the transition is indeed mean-field (see main text for further details). (Right panel) The critical exponents for the susceptibility ($\gamma$) and magnetization ($\beta$). We see that the high-temperature series expansion gives extremely precise mean-field results, indicating the lack of corrections to scaling behavior. We believe the deviations from mean-field predictions for the Monte Carlo are due to finite size effects, not all of which is taken into account by our correlation number scaling (see Sec. \ref{sec:monte_carlo} for more details).  Note that high-temperature expansion only gives $\gamma$ and not~$\beta$, for which only Monte Carlo data is provided.
	 }
\end{figure}

Fig. \ref{fig:mc_variation} shows the variation of the transition temperature, $T_c$ (in units of $J$), and the exponents for susceptibility and magnetization, $\gamma$ and $\beta$, for the different self-dual lattices (since we perform only high-temperature expansion for these lattices, we obtain only the susceptibility exponent). We see that as the coordination number is increased by 1, the $T_c$ also increases by approximately the same amount (within errors in estimating the critical temperature). Intuitively, this can be understood as a further indication of the mean-field nature of the phase-transition since in a mean-field setting, the transition temperature is proportional to the coordination number~\cite{Chaikin2000}: $T_c\propto r$.
Note that the critical exponents obtained by the series expansion are extremely close to the mean-field estimate of 1. This indicates that the series expansions are extremely well-behaved and there are no appreciable corrections to scaling. On the other hand, the Monte Carlo estimates are close to the mean-field value. We believe the deviations from mean-field predictions for the Monte Carlo are due to finite size effects, not all of which is taken into account by our correlation number scaling. The analysis of the Monte Carlo data is done as in Sec. \ref{sec:monte_carlo} and the detailed plots are not shown here for brevity.

Table ~\ref{tab:susseries} shows the results of the high-temperature series expansion for susceptibility for self-dual lattices $\{r,r\}$, $r = 5,6,7,8$. In addition, we have also provided the series for the $\{3,7\}$ and the $\{7,3\}$ lattices, the latter two dual of one-another.  The last two lattices will be needed when comparing our results to the Bethe lattice (see below). The exact values of $T_c$ and $\gamma$ obtained by the series expansions are shown in Table ~\ref{tab:susIDA}.

{
	\renewcommand{\arraystretch}{1.2}
	\begin{table}
		\centering
		\bgroup
		\def\arraystretch{1.5}%
		\resizebox{1\columnwidth}{!}{%
			\begin{tabular}{c r r r r r r}
				\hline
				\hline
				$n$ & $\{3,7\}$ & $\{7,3\}$ & $\{5,5\}$ & $\{6,6\}$ & $\{7,7\}$ & $\{8,8\}$ \\
				\hline
				1 & 3 & 7 & 5 & 6 & 7 & 8 \\
				2 & 6 & 42 & 20 & 30 & 42 & 56 \\
				3 & 12 & 238 & 80 & 150 & 252 & 392 \\
				4 & 24 & 1316 & 320 & 750 & 1512 & 2744 \\
				5 & 48 & 7196 & 1270 & 3750 & 9072 & 19208 \\
				6 & 96 & 39144 & 5040 & 18738 & 54432 & 134456 \\
				7 & 186 & 212394 & 20010 & 93630 & 326578 & 941192 \\
				8 & 360 & 1150968 & 79400 & 467862 & 1959384 & 6588328 \\
				9 & 702 & 6233150 & 315060 & 2337870 & 11755814 & 46118184 \\
				10 & 1368 & 33745698 & 1250260 & 11682090 & 70531944 & 322826520 \\
				11 & 2664 & 182669074 & 4961180 & 58374174 & 423174024 & 2259780264 \\
				12 & 5148 & 988735958 & 19686500 & 291689754 & 2538938220 & 15818424216 \\
				13 & 9948 & 5351558814 & 78119090 & 1457543742 & 15232993804 & 110728706088 \\
				14 & 19308 & 28964952422 & 309987000 & 7283195826 & 91394150092  & 775099098536 \\
				15 & 37434 & 156769556314 & 1230068820 &  & 548342025112 & 5425680781224 \\
				16 & 72504 & 848494238298 & 4881081760 &  & 3289914904549 & 37979675110288 \\
				17 & 140238 & & 19368790490 &  &  & 265857093267968 \\
				18 & 271242 & & & & & 1860995225360608 \\
				19 & 525528 & & & & & 13026935584994368 \\
				20 & 1017726 & & & & & \\
				21 & 1969458 & & & & & \\
				22 & 3811128 & & & & & \\
				23 & 7375278 & & & & & \\
				24 & 14279604 & & & & & \\
				\hline
				\hline
			\end{tabular}
		}
		\egroup
		\caption{The coefficients $x_n$ of the susceptibility series $\tilde{\chi} = 1 + \sum_{n=1}^{\infty} x_n\, v^n$. Our results for the high-temperature series expansion agree with \cite{Rietman1992} who obtained the series for~$\{5,5\}$ up to order~10 and for and for~$\{3,7\}$ up to order~11. }\label{tab:susseries}
	\end{table}
}

\begin{table}
	\centering
	\bgroup
	\def\arraystretch{1.5}%
	\begin{tabular}{c c c}
		\hline
		\hline
		lattice & $v_c$ & $\gamma$ \\
		\hline
		$\{3,7\}$ & $0.184764\pm 0.000004$ & $0.9999\pm 0.0004$ \\
		$\{7,3\}$ & $0.51\pm 0.04$ & $1.00\pm 0.02$ \\
		$\{5,5\}$ & $0.25200759 \pm 0.00000006$ & $1.000001 \pm 0.000005$ \\
		$\{6,6\}$ & $0.200125\pm 0.000001$ & $1.00002\pm 0.00004$ \\
		$\{7,7\}$ & $0.166673621\pm 4\cdot10^{-9}$ & $1.0000001\pm 2\cdot10^{-7}$ \\
		$\{8,8\}$ & $0.142857482725 \pm 7\cdot 10^{-12}$ & $0.9999999993 \pm 9\cdot 10^{-10}$ \\ 
		\hline
		\hline
	\end{tabular}
	\egroup

	\caption{Estimation of $v_c = \tanh(J/T_c)$ and the critical exponent~$\gamma$ obtained from the high-temperature series expansion of~$\tilde{\chi}$. The series analysis was done via first-order IDAs.
	}\label{tab:susIDA}
\end{table}

\subsection{Comparison to the Bethe lattice}

\begin{figure}
	\centering
	\includegraphics[width=0.5\columnwidth]{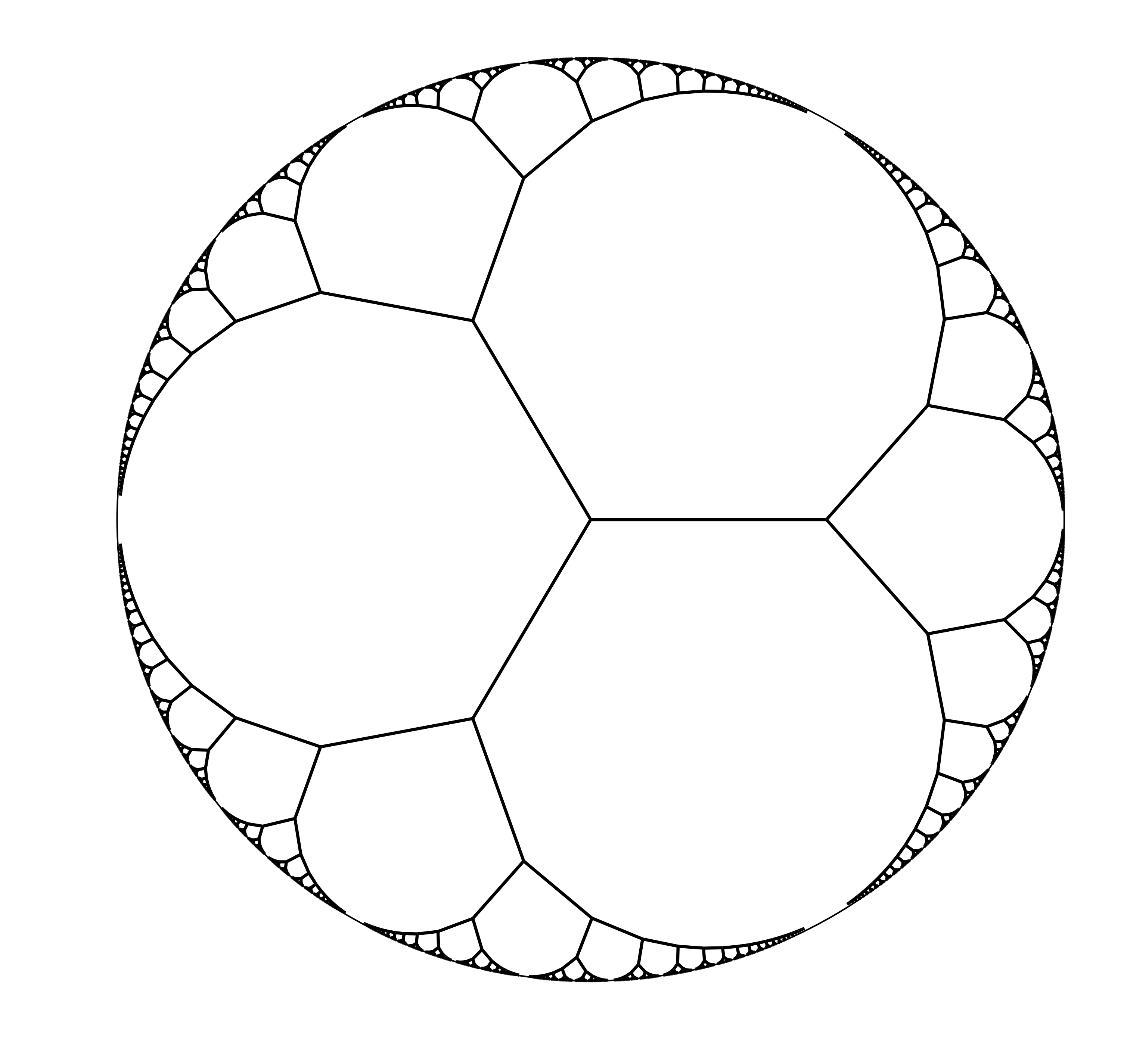}
	\caption{The Bethe lattice with coordination number $s=3$. It can be interpreted as a hyperbolic tiling with Schl\"afli symbol $\{\infty, s\}$.}
	\label{fig:bethe}
\end{figure}

In this section, we compare the critical properties of the above-analyzed hyperbolic lattices to the Bethe lattice. The lattice is the infinite $s$-regular tree, i.e. every vertex has $s$ neighbours and there are no cycles (closed loops) in the graph. It can be understood as a hyperbolic tiling where all faces have an infinite number of sides (see Fig.~\ref{fig:bethe}) which means that we can formally assign it the Schl\"afli symbol $\{\infty,s\}$.
Intuitively, for hyperbolic tilings $\{r,s\}$, where the number of edges around each face(~$r$) is large the Bethe lattice should be a good approximation.
Due to its tree-structure it is straightforward to solve the Ising model defined on the Bethe lattice~\cite{Baxter2013}.
The critical temperature is given by
\begin{align}\label{eqn:betheexacttemp}
T_c^\text{B} = \frac{2}{\ln \frac{s}{s-2}}
\end{align}
and hence
\begin{align}\label{eqn:betheexact}
v_c^\text{B} = \frac{1}{s-1}.
\end{align}

\begin{figure}
	\centering
	\includegraphics[width=0.49\columnwidth]{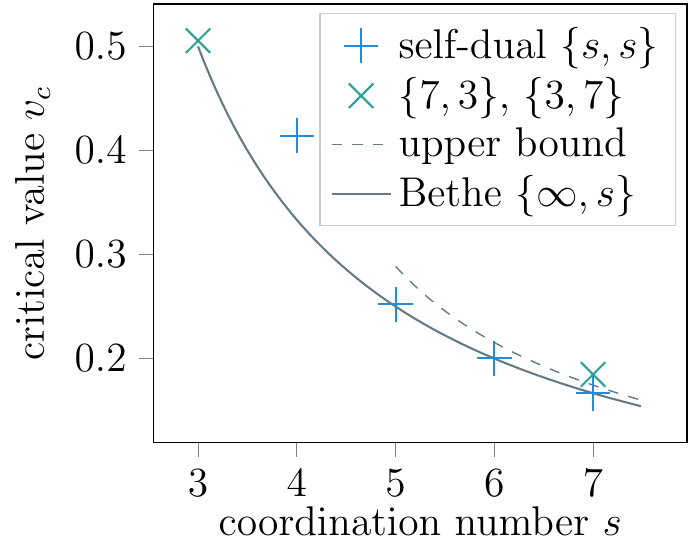}
	\hfil
	\includegraphics[width=0.49\columnwidth]{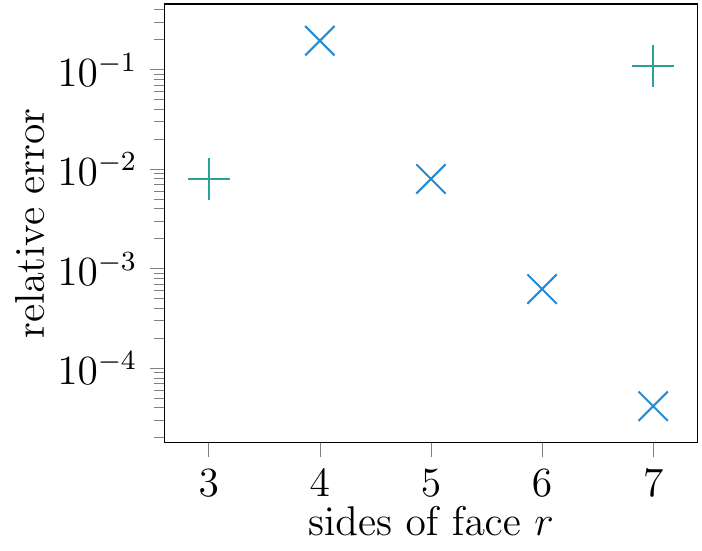}
	\caption{(a) Comparison of the results from Table ~\ref{tab:susIDA} to the exact solution of the Ising model on the Bethe lattice~$\{\infty,s\}$ given by Eq.~\eqref{eqn:betheexact}. The upper bound is obtained in~\cite{hypSAW} where the authors derive a bound on the coefficients of the series expansion of $\chi$ for self-dual lattices.  (b) The relative error when comparing the critical value $v_c$ of the hyperbolic Ising model and the Bethe lattice with the same coordination number. The  tessellations here are $\{3,7\}$, $\{4,4\}$ (euclidean square lattice), $\{5,5\}$, $\{6,6\}$ and $\{7,7\}$.}\label{fig:betheapprox}
\end{figure}

In Fig.~\ref{fig:betheapprox} the exact solution of the Bethe lattice is plotted together with the results of the high-temperature series expansion (see Table ~\ref{tab:susIDA}).
Evidently, Eq. ~\eqref{eqn:betheexact} provides a good approximation to the results that we obtained for the hyperbolic tilings.
The relative error between the critical values of the hyperbolic lattice and the Bethe lattice $|v_c - v_c^\text{B}| / v_c$ is shown in Fig.~\ref{fig:betheapprox}. The relative error decreases exponentially in the number of sides of a face~$r$. Note that the Ising model on 2D square-lattice ($\{4,4\}$ tiling) shows critical properties that are furthest from the mean-field behavior of the Bethe lattice.

\subsection{The intermediate phase and the second phase-transition}
\label{intermphase}
Recall that in addition to the phase-transition at $T_c$, the Ising model on the self-dual hyperbolic lattices undergoes a second phase-transition at a lower temperature~$\bar{T}_c$, where~$T_c, \bar{T}_c$ are related by the Kramers-Wannier duality relation [see discussion below Eq. \eqref{eqn:KWduality}]. Since the critical properties of the self-dual hyperbolic lattices are quite close to the Bethe lattice, in what follows, we describe the fate of this second phase-transition at $\bar{T}_c$ in the case of the Bethe lattice.

In the range of temperatures $\bar{T}_c< T < T_c$, the correlation function of two far away spins approaches $m^2$ instead of going to zero (that happens at temperatures $T>T_c$). Since the magnetization changes from zero to nonzero as $T$ changes from $T_c +\epsilon$ to $T_c - \epsilon$, $\epsilon\rightarrow0$, it is clear that the critical temperature $T^\text{B}_c$ in Eq.~\eqref{eqn:betheexacttemp} (obtained by analyzing the change in magnetization~\cite{Baxter2013}), corresponds to the critical temperature $T_c$. This implies that the low-temperature phase of the Bethe lattice Ising model, in fact, corresponds to the intermediate phase of the hyperbolic plane Ising models. This begs the question: what is $\bar{T}_c$ for the Bethe lattice when the system transitions to the pure ferromagnetic phase?

To answer this question, consider the correlation function of the Ising model on the Bethe lattice. It can be obtained exactly~\cite{kumar1976two} [cf.~Eq.~\eqref{eqn:high_temp_exp}]:
\begin{align}
\begin{split}
	\langle \sigma_{0}\, \sigma_{k} \rangle &= \frac{\sum_\sigma e^{-\beta H(\sigma)} \sigma_{0}\sigma_{k}}{\sum_\sigma e^{-\beta H(\sigma)}}\\
	&=  \frac{\sum_\sigma \prod_{(i,j)\in E} (1+\sigma_i \sigma_j v) \sigma_{0}\sigma_{k}}{\sum_\sigma \prod_{(i,j)\in E} (1+\sigma_i \sigma_j v)}
\end{split}
\end{align}

Observe that those terms in the product which contain an odd number of spin variables will cancel when summing over all spin configurations.
As taking the product can be interpreted as picking subsets of edges, we can interpret the numerator and the denominator as summing over subgraphs of the lattice (cf.~Appendix~\ref{sec:kramers_wannier}). Together with the previous observation we see that in the numerator the sum is taken over all subgraphs for which vertices 0 and $k$ have odd degree and all other vertices have even degree, while in the denominator all contributing graphs must have even degree.

For the Bethe lattice, the only contribution in the numerator is the line graph connecting $0$ and $k$. This graph has $\text{dist}(0,k)$ edges and hence, the numerator is exactly equal to $v^{\text{dist}(0,k)}$. The denominator has only the empty graph as a non-trivial contribution as the Bethe lattice has no finite subgraphs with all vertices of even degree.
Thus, for the Bethe lattice,
\begin{align}\label{eqn:bethe_correlation}
	\langle \sigma_{0} \sigma_{k} \rangle = v^{\text{dist}(0,k)}.
\end{align}
The correlation function is exponentially decaying {\em to zero} for $k \rightarrow \infty$ at any fixed non-zero temperature (just as for the hyperbolic Ising model in the intermediate phase) and is non-analytic at temperature~$0$ where it jumps to~$m^2 = 1$. Thus, for the Bethe lattice, $\bar{T}_c = 0$. Note that we assuming the convergence of the sum to arrive at this conclusion.

We tried to obtain signatures of this intermediate phase for the different hyperbolic lattices. However, with the system-sizes that we explored, the signal-to-noise ratio was not good enough to make any conclusive predictions. We hope to return to this problem in the future.

\section{The Ising model in 3D hyperbolic space}
\label{3DIsing}
So far, in this work, we analyzed the Ising model in 2D hyperbolic plane with Monte Carlo and high-temperature series expansion techniques.
While our analysis confirms the mean-field nature of the phase-transition, our results are not compatible with the conjectured formulas for critical exponents given by the field theory calculations~\cite{Mnasri2015}.
In this section, we analyze the Ising model in 3D hyperbolic space with periodic boundary condition, for which explicit $1/N$ computations revealed non-mean-field behavior.
For Monte Carlo simulations, we consider the $\{5,3,5\}$ lattice where the 3D hyperbolic space is tiled with dodecahedra in a way that there are 5 dodecahedra around each edge. The high-temperature analyses of the next section are done for $\{5,3,k\}$ lattices, where $k = 4,5,6$.

\subsection{Monte Carlo Analysis}
\label{3Dmc}
The Monte Carlo simulations were done for three different lattice sizes of the $\{5,3,5\}$ lattice with number of nodes in the vertices given by $N = 4428$, $14762$ and $390963$. We were able to perform simulations on only three lattice sizes since finding compactifications of 3D hyperbolic space is computationally even more challenging than for their 2D counterparts.

\begin{figure}
\centering
\includegraphics[width = 0.5\textwidth]{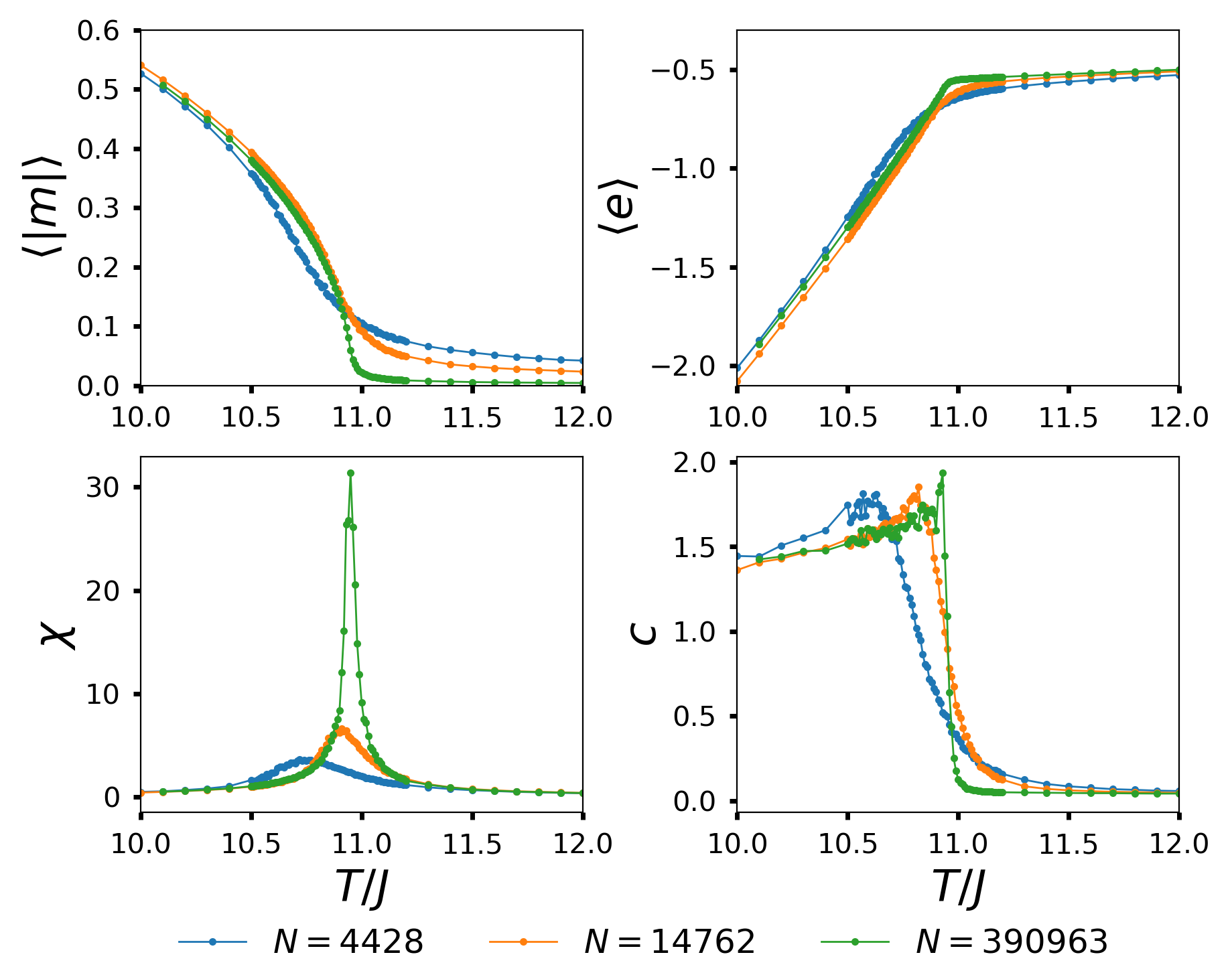}
\caption{\label{fig:mc_3D_1}Results of the Monte Carlo simulations for the hyperbolic space with $\{5,3,5\}$ tiling. The absolute magnetization per spin ($|m|$), the energy per spin ($\langle e\rangle$), average absolute susceptibility per spin ($\chi$) and the specific heat per spin ($c$) are plotted  in the top left, top right, bottom left and bottom right panels respectively.}
\end{figure}

The results of the Monte Carlo simulations for the absolute magnetization per spin, the energy per spin, average absolute susceptibility per spin and the specific heat per spin are shown in Fig. \ref{fig:mc_3D_1}. The smaller two lattices were equilibrated with $10^4$ sweeps of the lattice and the Monte Carlo measurements were done over $10^5$ sweeps of the lattice. The finite size scaling analysis was done as in Sec. \ref{sec:monte_carlo} using the correlation number exponent $\mu = 2$, thereby avoiding difficulties associated with the multiple linear dimensions. Since the connectivity of the 3D lattice is higher (in the case considered, the number of nearest neighbors is 12 for any given spin), we expect the correlation number scaling to yield better results than that obtained for the 2D lattices analyzed in Sec. \ref{sec:variation_with_curvature}.

From the finite size scaling analysis, we infer that the critical temperature is $T_c = 10.96\pm 0.01$ and $\gamma = 0.97\pm 0.02$, $\beta = 0.51\pm 0.04$, which are close to the mean-field predictions. Comparing our results to those obtained by field theory ($1/N$)  computations~\cite{Mnasri2015}, we see that the susceptibility exponent is not compatible with the field theory computations, who obtain $\gamma = 2$. On the other hand, the magnetization exponent, inferred from scaling relations, together with the field theory computations, agree. As in the 2D case, the peak in the specific heat did not develop upon increase of system size, which indicates that the specific heat does not diverge and we expect $\alpha = 0$ in this case as well.

\begin{figure}
\centering
\includegraphics[width = 0.5\textwidth]{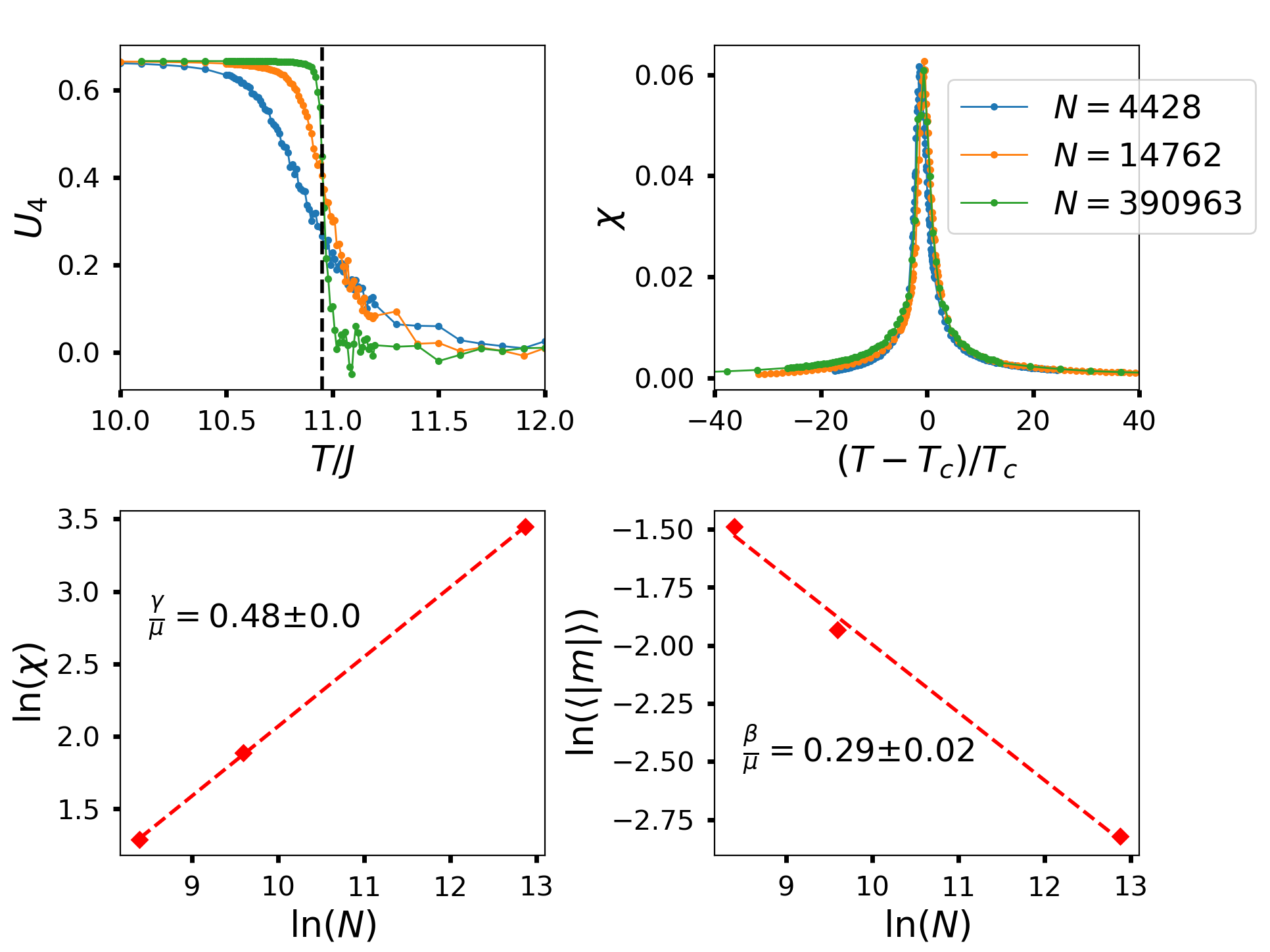}
\caption{\label{fig:mc_3D_2} Results of the Monte Carlo simulations: (top left) the fourth Binder cumulant, (top right) data collapse for average absolute susceptibility and the linear fits are system size scaling for average absolute magnetization (bottom left) and average absolute susceptibility per spin (bottom right). The vertical line in the top left panel at $T/J = 10.96$ (see main text for error estimate) indicates the location of phase-transition temperature obtained from the data collapse. The linear fits in the bottom panels denote are obtained from the finite size analysis with respect to the number of spins. The slope for the linear fit for average absolute susceptibility (magnetization) per spin provides the ratio $\gamma/\mu$ ($\beta/\mu$), where $\gamma$ ($\beta$) denote the exponents for susceptibility (magnetization) and $\mu$ is the scaling of the coherence number (see Sec. \ref{sec:monte_carlo} for details). In the bottom panels, the error indicated is only the fit error, the actual value and the error in the exponent is provided in the main text. }
\end{figure}

\subsection{High-Temperature Series Analysis}
\label{3dtemp}

To verify the results obtained by Monte Carlo simulation, we also compute the high-temperature series expansion of the susceptibility for both the $\{5, 3, 5\}$ and $\{5, 3, 4\}$ lattices.
We also consider the  $\{5, 3, 6\}$ lattice which is a 3D generalization of the dual of the Bethe lattice.
The Bethe lattice can be interpreted as a tessellation where the faces are $\infty$-gons, i.e. they have the (only possible) tessellation of the infinite line~$\mathbb{R}$ at their boundary.
The $\{5, 3, 6\}$ tessellation is dual to the $\{6, 3, 5\}$ tessellation which is a space tessellated by non-compact polyhedra which have a hexagonal tiling~$\{6, 3\}$ at their boundary.

The first four graphs contributing to those series, together with their embeddig numbers per site,  are given in Tab.~\ref{fig:3d-series}.
The weights $W(g)$ are the same as in Sec.~\ref{sec:series_expansion}.
Summing the contributions yields the inverse susceptibility $\bar\chi^{-1}$.
Inverting that series gives the coefficients in Tab.~\ref{tab:susseries3d}.
For analysis of the series, we use FO-IDAs. Averaging over 8 different approximants using a minimum number of 8 terms yields an estimate of the critical properties tabulated in Tab.~\ref{tab:susIDA3d}.
The results are very close to the mean field predictions and the $\{5, 3, 5\}$ result is close to the result of the Monte~Carlo simulation. The exponents $\gamma$ are not compatible with the field-theory~($1/N$) prediction of $\gamma=2$~\cite{Mnasri2015}.

\begin{table}
\centering
\bgroup
\def\arraystretch{1.5}%
\begin{tabular}{c c c c c}
\hline
\hline
$g$ &
\begin{minipage}[c]{1.5cm}\vspace{.2cm}\includegraphics[width=1.3cm]{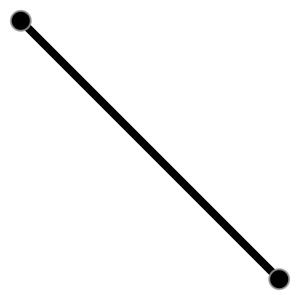}\end{minipage}&
\begin{minipage}[c]{1.5cm}\vspace{.2cm}\includegraphics[width=1.3cm]{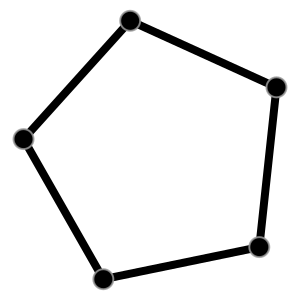}\end{minipage}&
\begin{minipage}[c]{1.5cm}\vspace{.2cm}\includegraphics[width=1.3cm]{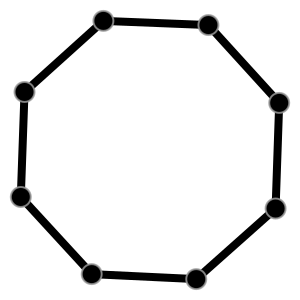}\end{minipage}&
\begin{minipage}[c]{1.5cm}\vspace{.2cm}\includegraphics[width=1.3cm]{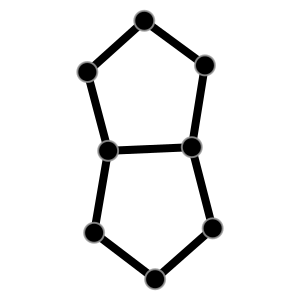}\end{minipage}\\[0.7cm]
\hline
$c_{\{5, 3, 4\}}(g)$ & 6 & 24/5 & 36 & 36\\[0.1cm]
$c_{\{5, 3, 5\}}(g)$ & 6 & 6 & 60 & 60\\[0.1cm]
$c_{\{5, 3, 6\}}(g)$ & 6 & 36/5 & 90 & 90\\[0.1cm]
\hline
\hline
\end{tabular}
\egroup
\caption{\label{fig:3d-series}The four smallest graphs (exlcuding the single bond) contributing to the susceptibility series of the Ising model in hyperbolic space with $\{5, 3, 4\}$, $\{5, 3, 5\}$ and $\{5, 3, 6\}$ tiling. The number of embeddings per site $c(g)$ for each graph~$g$ is given explicitly for all three tilings}
\end{table}

\begin{table}
	\centering
	\bgroup
	\def\arraystretch{1.5}%
		\begin{tabular}{c r r r }
			\hline
			\hline
			$n$ & $\{5, 3, 4\}$ & $\{5, 3, 5\}$ & $\{5, 3, 6\}$\\
			\hline
			1 & 12 & 12 & 12\\
			2 & 132 & 132 & 132\\
			3 & 1452 & 1452 & 1452\\
			4 & 15972 & 15972 & 15792\\
			5 & 175644 & 175632 & 175620\\
			6 & 1931556 & 1931292 & 1931028\\
			7 & 21241356 & 21237012 & 21232668\\
			8 & 233590980 & 233526972 & 233462868\\
			9 & 2568797772 & 2567915412 & 2567030988\\
			10 & 28249045956 & 28237381152 & 28225683060\\
			\hline
			\hline
		\end{tabular}
		\egroup
	\caption{The coefficients $x_n$ of the susceptibility series $\tilde{\chi} = 1 + \sum_{n=1}^{\infty} x_n\, v^n$. For the Ising model on hyperbolic tilings in three dimensions}\label{tab:susseries3d}
\end{table}

\begin{table}
	\centering
	\bgroup
	\def\arraystretch{1.5}%
	\begin{tabular}{c c c}
		\hline
		\hline
		lattice & $v_c$ & $\gamma$ \\
		\hline
		$\{5,3,4\}$ & $0.09093417 \pm 0.00000008$ & $1.000012 \pm 0.000003$ \\
		$\{5,3,5\}$ & $0.09094066 \pm 0.00000013$ & $1.000023 \pm 0.000005$ \\
		$\{5,3,6\}$ & $0.09094723 \pm 0.00000018$ & $1.000037 \pm 0.000007$ \\
		\hline
		\hline
	\end{tabular}
	\egroup
	\caption{\label{tab:susIDA3d}Estimation of $v_c = \tanh(J/T_c)$ and the critical exponent~$\gamma$ obtained from the high-temperature series expansion of~$\tilde{\chi}$ for hyperbolic tilings in three dimensions. The series analysis was done via first-order IDAs.
	}
\end{table}

\section{Conclusion and Perspectives}
\label{sec:conclusion}
To summarize, we have analyzed the critical properties of the Ising model in hyperbolic space using Monte Carlo and series expansion methods. The negative curvature of hyperbolic space leads to a comparable number of spins on the boundary as in the bulk of an open hyperbolic space, which leads to large boundary effects. While these boundary effects can be interesting in themselves, they obscure the bulk properties of the model. We analyze the bulk properties of the Ising model in hyperbolic spaces with periodic boundary condition. First, we find compactifications of the hyperbolic manifolds, which are manifolds with genus larger than 1, in contrast to the euclidean plane with periodic boundary condition, a torus, with genus 1. Performing Monte Carlo simulations on these compactified hyperbolic manifolds, we infer the critical temperature and critical exponents for several two and three dimensional lattices. We obtain results that are close to mean-field predictions. Subsequently, we perform high-temperature series analysis for the different lattices which confirm the mean-field nature of the critical behavior and are close to our Monte Carlo findings. We analyze the variation of the critical temperature as a function of curvature and explain how the properties of the Ising model on the Bethe lattice can be viewed as an asymptotic case of that on the different hyperbolic lattices. Recently, 2D mesoscopic superconducting circuit lattices have been engineered which implement spin-systems in hyperbolic lattices~\cite{Kollar2019}.  With this remarkable experimental progress, we are optimistic that statistical mechanical models in hyperbolic space will be experimentally realized in the near future.

Before concluding, we outline several possible research directions that are of interest.
First, the 2D Ising model on the hyperbolic plane with free boundary conditions is expected to exhibit an intermediate phase, between the ferromagnetic and the disordered phases. This intermediate phase, absent in the euclidean space Ising model, shows broken translational invariance, where the lattice is covered with infinitely large, infinitely many magnetized domains. We were unable to find conclusive evidence of this phase in our Monte Carlo simulations for the case of periodic boundary conditions. This could be due to the fact that the sizes of the systems analyzed were not large enough to ensure several magnetized clusters to form in addition to domains of randomized spins. We emphasize that the main limitation is not the Monte Carlo aspect of the simulation, but finding the compactifcations of the hyperbolic space itself, which is computationally costly. Larger scale Monte Carlo simulations on compactified hyperbolic planes may be able to find evidence for this intermediate phase.
Second, the found compactifications of hyperbolic space are valuable for analyzing bulk properties of different gauge and matter spin-systems in these spaces --- an interesting problem, much less explored than its flat-space counterpart.
Third, the high genus compactified hyperbolic manifolds can be used to implement quantum codes such as toric codes, which are promising for quantum information processing. In contrast to their euclidean space counterparts, for toric codes on these manifolds, the number of encoded logical qubits scales proportionally with the number of actual physical qubits \cite{Freedman2002, Breuckmann2017} .
Interestingly, the decoding of these hyperbolic space toric codes can be related to the random-bond Ising model in hyperbolic space \cite{dklp}, which, to the best of our knowledge, has not been analyzed before. The ferromagnetic to paramagnetic phase-transition points provide the threshold for successful decoding of the error syndrome for the different decoders of the quantum code. We hope to report on properties of the random-bond Ising model on hyperbolic manifolds in the near future.

\section*{Acknowledgments}
We thank Leonid Pryadko for encouraging discussions.
N.P.B. is supported by the UCLQ Fellowship.
A.R. acknowledges the support of the Alexander von Humboldt foundation.

\appendix
\section{Regular Tilings of the Hyperbolic Plane with Periodic Boundary Conditions}
In this appendix, we provide some technical details for finding regular tessellations of the hyperbolic plane~$\mathbb{H}^2$ with periodic boundary condition.
\label{hyperbolic_tiling}
	\subsection{Regular Tilings of $\mathbb{H}^2$}
	A tiling is \textit{regular} if its faces are unilateral, equiangular and identical and the same number of faces meet at every vertex.
	Regular tilings are classified by their \textit{Schl\"afli symbol} $\{r,s\}$ where $r$ stands for the number of edges in a face and $s$ stands for the number of faces meeting at a vertex.
	Familiar examples of regular tilings are the square tiling~$\{4,4\}$, the hexagonal tiling~$\{6,3\}$ and the triangular tiling~$\{3,6\}$.
	These are in fact all possible regular tilings in the euclidean plane.
	The hyperbolic plane admits an infinite number of different regular tilings~$\{r,s\}$ where $r$ and $s$ can be any integers~$\geq 3$ satisfying $1/r + 1/s < 1/2$.

	\subsection{Periodic Boundaries}
	To analyze the Ising model with Monte Carlo and in particular with finite size scaling it is necessary to have a family of surfaces of growing area (and with the same tessellation).
	For the Ising model in euclidean space we can consider the tessellation of square patches of increasing size with boundaries.
	The values of the spins at the boundaries can either be fixed to a value (fixed boundaries) or by treating them the same as the bulk spins (free boundaries).
	Alternatively, we can identify the two pairs of opposing boundaries of each of the patches, which effectively creates a family of tori of increasing area.
	In euclidian space, both of these approaches are valid to perform a finite size scaling analysis as in the limit of infinite system size the effects of the boundaries will vanish.

	However, this is not true in hyperbolic space.
	It can be shown that the number of vertices~$N_\partial$ at the boundary of a tessellated patch in hyperbolic space is a constant fraction of the total number of vertices $N_{\partial} = C\, N$.
	The asymptotic value for $C$ can be derived using a recursion relation~\cite{moran}.
	It is given by
	\begin{align}
		C = \lim_{N \rightarrow \infty} \frac{N_{\partial}}{N} = 1-\frac{2}{\lambda+\sqrt{\lambda^2 + 4}},
	\end{align}
	where $\lambda = rs-2r-2s+2$.
	Note that for all possible hyperbolic tilings, where $1/r + 1/s < 1/2$, we have that $N_{\partial} / N > 1/2$.
	For the Ising model this means that asymptotically more than half of spins will participate in interaction terms which are different from the bulk spins.
	Hence it is expected that boundaries will change the behavior of the model in a significant way, regardless of whether the boundary conditions are free or fixed.
	We therefore want to stress that taking open boundary surfaces of increasing size is not the correct limit to take for finite size scaling analysis.
	This has been observed in \cite{hasegawa} where the authors try to mitigate this effect by ignoring spins close to the boundary (cf. Sec.~\ref{sec:monte_carlo}).

	We avoid this problem by performing finite size scaling on families of closed surfaces of increasing area.
	The process of producing a finite surface is not as trivial as for euclidean tilings, where it suffices to identify two pairs of opposite boundaries.
	The reason for this is that translations in curved spaces do not commute.
	This is well known for translations for the sphere and in fact it can be taken as an alternative definition of curvature.
	The details of this construction can be found in \cite{hyperbolic_constr_thresh}.
	For each $\{r,s\}$-tiling of the hyperbolic plane we obtain a family of closed surfaces of growing area with the same $\{r,s\}$-tiling.

	These surfaces will necessarily have non-trivial topology due to the Gau\ss -Bonnet theorem, which for negatively-curved, orientable surfaces states that the surface area is proportional to the genus (the number of handles).
	This is not a cause for concern as the largest region in which the lattice is identical to the infinite lattice grows with its total area.

\section{Derivation of the High-Temperature Susceptibility Series}\label{ap:series_derivation}
The susceptibility is given by
\begin{align}
\chi = \lim_{h \rightarrow 0}\, \frac{\partial m}{\partial h} = \beta\, \frac{1}{N}\sum_{i,j = 1}^{N} \langle \sigma_i\, \sigma_j \rangle\, -\, \beta\, m^2
\end{align}
where $m = \beta^{-1}\, \partial f / \partial h$ is the average magnetization per spin.
We will perform an expansion in the quantity
\begin{align}
\tilde{\chi} =  \frac{1}{N}\sum_{i,j = 1}^{N} \langle \sigma_i\, \sigma_j \rangle .
\end{align}

It was shown in \cite{singh87} that all non-zero contributions in the high-temperature expansion of the \emph{inverse} of the sum over two-point correlators~$\tilde{\chi}^{-1}$ are coming from biconnected graphs.
A \textit{biconnected graph} is a connected graph which stays connected if any of its vertices and all edges connected to it are being removed.\footnote{Biconnected graphs are sometimes also referred to as ``star-graphs'' in the literature.}
The restriction to biconnectedness drastically reduces the number of graphs that we have to generate when compared to the naive expansion.

The expansion of $\tilde{\chi}^{-1}$ is performed as follows:
Let us define a matrix $M$ which contains the values of every two-point correlator, i.e.~$M_{i,j} = \langle \sigma_i\, \sigma_j \rangle$. An elementary calculation shows that
\begin{align}\label{eqn:invsus}
N\, \tilde{\chi}^{-1} = \sum_{i,j = 1}^{N} M^+_{i,j}
\end{align}
where $M^+$ is the Moore-Penrose pseudo-inverse.
For an Ising model on a graph $G$ with vertex-set~$V_G$ we define the \emph{amplitude}
\begin{align}\label{eqn:susamplitudesdef}
\psi(G) = \sum_{i,j\in V_G} M^+_{i,j}\, -\, |V_G|.
\end{align}
Let $B(G)$ be the set of biconnected subgraphs of~$G$ induced by a subset of the edges of~$G$, $R(G)$ the set of representatives of each isomorphism-class of~$B(G)$ and~$c(g)$ the number of embeddings of a subgraph $g \subset G$ divided by the number of vertices of $G$. It is shown in~\cite{singh87} that for any graph~$G$ there exists a function $W : B(G) \rightarrow \mathbb{R}$ assigning each biconnected subgraph $g$ of~$G$ a weight such that
\begin{align}
\psi(G) = \sum_{g \in B(G)} W(g) \label{eqn:susamplitudesweights} = \sum_{g \in R(G)} c'(g)\, W(g)
\end{align}
where $c'(g)$ is the embedding constant of $g$ in $G$.
Combining Eqs. \ref{eqn:invsus}, \ref{eqn:susamplitudesdef} and \ref{eqn:susamplitudesweights} and applying the result to the tiling of the infinite plane we obtain
\begin{align}\label{eqn:sus_seriesA}
\tilde{\chi}^{-1} = 1\, + \, \sum_{g}\, c(g)\, W(g)
\end{align}
where each $g$ is a representative of a biconnected subgraph of the infinite tessellation of $\mathbb{H}^2$.
We can now obtain the expansion of $\tilde{\chi}^{-1}$ to any desired order by generating biconnected subgraphs $g$ of the tessellation, their embedding constants $c(g)$ and determining their weights~$W(g)$ using Eqs.~\ref{eqn:susamplitudesdef} and~\ref{eqn:susamplitudesweights}.

\section{Kramers-Wannier Duality}
\label{sec:kramers_wannier}
	For a finite system with $N$ spins and no external magnetic field ($h=0$) we can rewrite the partition function~$Z_N$ in two different ways.
	We assume for our proof that the system is defined on a self-dual tiling of a closed surface.

	\emph{High-Temperature Expansion:}
		Let $Z_1$ be the set of subsets $\gamma \subset E$ such that in the subgraph induced by one of its elements every vertex has even degree.
		The partition function can be rewritten as follows.
		\begin{align}\label{eqn:high_temp_exp}
		\begin{split}
		\mathcal{Z}(K)    &= \sum_{{\sigma}\in \{\pm 1\}^N} \prod_{(i,j)\in E} \exp(K \sigma_i \sigma_j)\\
		&= (\cosh K)^{|E|} \sum_{{\sigma}} \prod_{(i,j)} (1+ \sigma_i \sigma_j \tanh K)\\
		&= 2^N (\cosh K)^{|E|} \sum_{\gamma \in Z_1} (\tanh K)^{|\gamma|}
		\end{split}
		\end{align}
		In the last equation we have used that the expansion of the product $\prod_{(i,j)\in E} (1+ \sigma_i \sigma_j \tanh K)$ gives a sum over all subsets of edges $\gamma \subset E$ where each term is of the form $(\tanh K)^{|\gamma |} \prod_{(i,j)\in S} \sigma_i \sigma_j$.
		All terms where the subgraph induced by $\gamma$ has a vertex $v$ with odd degree have to cancel as states~${\sigma}$ with~$\sigma_v = \pm 1$ appear in the sum.
		Hence only terms for which~$\gamma$ induces an even degree subgraph give a non-zero contribution.
		These are exactly the elements of~$Z_1$.

	 	\emph{Low-Temperature Expansion:}
		Let $B^1$ be the set of all subsets of edges $\bar{\gamma} \subset E$ such that every face is surrounded by an even number of edges in $\bar{\gamma}$.
		\begin{align}\label{eqn:low_temp_exp}
		\begin{split}
		\mathcal{Z}(\bar{K})    &= \sum_{{\sigma}\in \{\pm 1\}^N} \prod_{(i,j)\in E} \exp(\bar{K} \sigma_i \sigma_j)\\
		&= 2 \sum_{\bar{\gamma} \in B^1} \exp(\bar{K})^{|E|-2|\bar{\gamma}|}\\
		&= 2 \exp(\bar{K})^{|E|} \sum_{\bar{\gamma} \in B^1} \exp(-2 \bar{K})^{|\bar{\gamma}|}
		\end{split}
		\end{align}
		Note that the first equality directly follows from the definition of $B^1$.

	For all orders smaller than the length of a non-contractible loop there exists a 1-to-1 mapping from~$B^1$ to~$Z_1$.
	In this case we can choose
	\begin{align}\label{eqn:dual_temp}
		\exp(-2\bar{K}) = \tanh(K)
	\end{align}
	and we see that the expressions given in Eqs.~\ref{eqn:high_temp_exp} and~\ref{eqn:low_temp_exp} are proportional.
	This gives Eq.~\ref{eqn:KWduality}.

	Under the assumption that the phase-transition point is unique, as it is the case for the euclidean Ising model, this suffices to determine the critical temperature by solving for $K = \bar{K}$ in \eqref{eqn:dual_temp}.
	However, we know that there exist two different critical temperatures in the hyperbolic Ising model~\cite{Wu1996,Wu2000} and Equation~\ref{eqn:KWduality} allows us to determine the dual temperature.

\bibliography{library_1}

\begin{thebibliography}{38}%
\makeatletter
\providecommand \@ifxundefined [1]{%
 \@ifx{#1\undefined}
}%
\providecommand \@ifnum [1]{%
 \ifnum #1\expandafter \@firstoftwo
 \else \expandafter \@secondoftwo
 \fi
}%
\providecommand \@ifx [1]{%
 \ifx #1\expandafter \@firstoftwo
 \else \expandafter \@secondoftwo
 \fi
}%
\providecommand \natexlab [1]{#1}%
\providecommand \enquote  [1]{``#1''}%
\providecommand \bibnamefont  [1]{#1}%
\providecommand \bibfnamefont [1]{#1}%
\providecommand \citenamefont [1]{#1}%
\providecommand \href@noop [0]{\@secondoftwo}%
\providecommand \href [0]{\begingroup \@sanitize@url \@href}%
\providecommand \@href[1]{\@@startlink{#1}\@@href}%
\providecommand \@@href[1]{\endgroup#1\@@endlink}%
\providecommand \@sanitize@url [0]{\catcode `\\12\catcode `\$12\catcode
  `\&12\catcode `\#12\catcode `\^12\catcode `\_12\catcode `\%12\relax}%
\providecommand \@@startlink[1]{}%
\providecommand \@@endlink[0]{}%
\providecommand \url  [0]{\begingroup\@sanitize@url \@url }%
\providecommand \@url [1]{\endgroup\@href {#1}{\urlprefix }}%
\providecommand \urlprefix  [0]{URL }%
\providecommand \Eprint [0]{\href }%
\providecommand \doibase [0]{http://dx.doi.org/}%
\providecommand \selectlanguage [0]{\@gobble}%
\providecommand \bibinfo  [0]{\@secondoftwo}%
\providecommand \bibfield  [0]{\@secondoftwo}%
\providecommand \translation [1]{[#1]}%
\providecommand \BibitemOpen [0]{}%
\providecommand \bibitemStop [0]{}%
\providecommand \bibitemNoStop [0]{.\EOS\space}%
\providecommand \EOS [0]{\spacefactor3000\relax}%
\providecommand \BibitemShut  [1]{\csname bibitem#1\endcsname}%
\let\auto@bib@innerbib\@empty
\bibitem [{\citenamefont {Callan}\ and\ \citenamefont
  {Wilczek}(1990)}]{Callan1990}%
  \BibitemOpen
  \bibfield  {author} {\bibinfo {author} {\bibfnamefont {C.~G.}\ \bibnamefont
  {Callan}}\ and\ \bibinfo {author} {\bibfnamefont {F.}~\bibnamefont
  {Wilczek}},\ }\href {\doibase https://doi.org/10.1016/0550-3213(90)90451-I}
  {\bibfield  {journal} {\bibinfo  {journal} {Nuclear Physics B}\ }\textbf
  {\bibinfo {volume} {340}},\ \bibinfo {pages} {366 } (\bibinfo {year}
  {1990})}\BibitemShut {NoStop}%
\bibitem [{\citenamefont {Wald}(1994)}]{Wald1994}%
  \BibitemOpen
  \bibfield  {author} {\bibinfo {author} {\bibfnamefont {R.}~\bibnamefont
  {Wald}},\ }\href {https://books.google.de/books?id=Iud7eyDxT1AC} {\emph
  {\bibinfo {title} {Quantum Field Theory in Curved Spacetime and Black Hole
  Thermodynamics}}},\ Chicago Lectures in Physics\ (\bibinfo  {publisher}
  {University of Chicago Press},\ \bibinfo {year} {1994})\BibitemShut {NoStop}%
\bibitem [{\citenamefont {{Kl\'eman, M.}}(1982)}]{Kleman1982}%
  \BibitemOpen
  \bibfield  {author} {\bibinfo {author} {\bibnamefont {{Kl\'eman, M.}}},\
  }\href {\doibase 10.1051/jphys:019820043090138900} {\bibfield  {journal}
  {\bibinfo  {journal} {J. Phys. France}\ }\textbf {\bibinfo {volume} {43}},\
  \bibinfo {pages} {1389} (\bibinfo {year} {1982})}\BibitemShut {NoStop}%
\bibitem [{\citenamefont {Rubinstein}\ and\ \citenamefont
  {Nelson}(1983)}]{Rubinstein1983}%
  \BibitemOpen
  \bibfield  {author} {\bibinfo {author} {\bibfnamefont {M.}~\bibnamefont
  {Rubinstein}}\ and\ \bibinfo {author} {\bibfnamefont {D.~R.}\ \bibnamefont
  {Nelson}},\ }\href {\doibase 10.1103/PhysRevB.28.6377} {\bibfield  {journal}
  {\bibinfo  {journal} {Phys. Rev. B}\ }\textbf {\bibinfo {volume} {28}},\
  \bibinfo {pages} {6377} (\bibinfo {year} {1983})}\BibitemShut {NoStop}%
\bibitem [{\citenamefont {Nelson}(2002)}]{Nelson2002}%
  \BibitemOpen
  \bibfield  {author} {\bibinfo {author} {\bibfnamefont {D.}~\bibnamefont
  {Nelson}},\ }\href {https://books.google.de/books?id=CGdgQgAACAAJ} {\emph
  {\bibinfo {title} {Defects and Geometry in Condensed Matter Physics}}}\
  (\bibinfo  {publisher} {Cambridge University Press},\ \bibinfo {year}
  {2002})\BibitemShut {NoStop}%
\bibitem [{\citenamefont {Freedman}\ \emph {et~al.}(2002)\citenamefont
  {Freedman}, \citenamefont {Meyer},\ and\ \citenamefont {Luo}}]{Freedman2002}%
  \BibitemOpen
  \bibfield  {author} {\bibinfo {author} {\bibfnamefont {M.}~\bibnamefont
  {Freedman}}, \bibinfo {author} {\bibfnamefont {D.}~\bibnamefont {Meyer}}, \
  and\ \bibinfo {author} {\bibfnamefont {F.}~\bibnamefont {Luo}},\ }\enquote
  {\bibinfo {title} {Z2-systolic freedom and quantum codes},}\ \ (\bibinfo
  {year} {2002})\ pp.\ \bibinfo {pages} {287--320}\BibitemShut {NoStop}%
\bibitem [{\citenamefont {Breuckmann}\ \emph {et~al.}(2017)\citenamefont
  {Breuckmann}, \citenamefont {Vuillot}, \citenamefont {Campbell},
  \citenamefont {Krishna},\ and\ \citenamefont {Terhal}}]{Breuckmann2017}%
  \BibitemOpen
  \bibfield  {author} {\bibinfo {author} {\bibfnamefont {N.~P.}\ \bibnamefont
  {Breuckmann}}, \bibinfo {author} {\bibfnamefont {C.}~\bibnamefont {Vuillot}},
  \bibinfo {author} {\bibfnamefont {E.}~\bibnamefont {Campbell}}, \bibinfo
  {author} {\bibfnamefont {A.}~\bibnamefont {Krishna}}, \ and\ \bibinfo
  {author} {\bibfnamefont {B.~M.}\ \bibnamefont {Terhal}},\ }\href {\doibase
  10.1088/2058-9565/aa7d3b} {\bibfield  {journal} {\bibinfo  {journal} {Quantum
  Science and Technology}\ }\textbf {\bibinfo {volume} {2}},\ \bibinfo {pages}
  {035007} (\bibinfo {year} {2017})}\BibitemShut {NoStop}%
\bibitem [{\citenamefont {Mnasri}\ \emph {et~al.}(2015)\citenamefont {Mnasri},
  \citenamefont {Jeevanesan},\ and\ \citenamefont {Schmalian}}]{Mnasri2015}%
  \BibitemOpen
  \bibfield  {author} {\bibinfo {author} {\bibfnamefont {K.}~\bibnamefont
  {Mnasri}}, \bibinfo {author} {\bibfnamefont {B.}~\bibnamefont {Jeevanesan}},
  \ and\ \bibinfo {author} {\bibfnamefont {J.}~\bibnamefont {Schmalian}},\
  }\href {\doibase 10.1103/PhysRevB.92.134423} {\bibfield  {journal} {\bibinfo
  {journal} {Phys. Rev. B}\ }\textbf {\bibinfo {volume} {92}},\ \bibinfo
  {pages} {134423} (\bibinfo {year} {2015})}\BibitemShut {NoStop}%
\bibitem [{\citenamefont {Benjamini}\ \emph {et~al.}(1999)\citenamefont
  {Benjamini}, \citenamefont {Lyons}, \citenamefont {Peres},\ and\
  \citenamefont {Schramm}}]{Benjamini1999}%
  \BibitemOpen
  \bibfield  {author} {\bibinfo {author} {\bibfnamefont {I.}~\bibnamefont
  {Benjamini}}, \bibinfo {author} {\bibfnamefont {R.}~\bibnamefont {Lyons}},
  \bibinfo {author} {\bibfnamefont {Y.}~\bibnamefont {Peres}}, \ and\ \bibinfo
  {author} {\bibfnamefont {O.}~\bibnamefont {Schramm}},\ }\href {\doibase
  10.1214/aop/1022677450} {\bibfield  {journal} {\bibinfo  {journal} {Ann.
  Probab.}\ }\textbf {\bibinfo {volume} {27}},\ \bibinfo {pages} {1347}
  (\bibinfo {year} {1999})}\BibitemShut {NoStop}%
\bibitem [{\citenamefont {{Benjamini}}\ and\ \citenamefont
  {{Schramm}}(2001)}]{Benjamini2001}%
  \BibitemOpen
  \bibfield  {author} {\bibinfo {author} {\bibfnamefont {I.}~\bibnamefont
  {{Benjamini}}}\ and\ \bibinfo {author} {\bibfnamefont {O.}~\bibnamefont
  {{Schramm}}},\ }\href@noop {} {\bibfield  {journal} {\bibinfo  {journal} {J.
  Amer. Math. Soc.}\ }\textbf {\bibinfo {volume} {14}},\ \bibinfo {eid}
  {math/9912233} (\bibinfo {year} {2001})},\ \Eprint
  {http://arxiv.org/abs/math/9912233} {arXiv:math/9912233 [math.PR]}
  \BibitemShut {NoStop}%
\bibitem [{\citenamefont {Baek}\ \emph {et~al.}(2009)\citenamefont {Baek},
  \citenamefont {Minnhagen},\ and\ \citenamefont {Kim}}]{Baek2009}%
  \BibitemOpen
  \bibfield  {author} {\bibinfo {author} {\bibfnamefont {S.~K.}\ \bibnamefont
  {Baek}}, \bibinfo {author} {\bibfnamefont {P.}~\bibnamefont {Minnhagen}}, \
  and\ \bibinfo {author} {\bibfnamefont {B.~J.}\ \bibnamefont {Kim}},\ }\href
  {\doibase 10.1103/PhysRevE.79.011124} {\bibfield  {journal} {\bibinfo
  {journal} {Phys. Rev. E}\ }\textbf {\bibinfo {volume} {79}},\ \bibinfo
  {pages} {011124} (\bibinfo {year} {2009})}\BibitemShut {NoStop}%
\bibitem [{\citenamefont {Wu}(1996)}]{Wu1996}%
  \BibitemOpen
  \bibfield  {author} {\bibinfo {author} {\bibfnamefont {C.~C.}\ \bibnamefont
  {Wu}},\ }\href {\doibase 10.1007/BF02175564} {\bibfield  {journal} {\bibinfo
  {journal} {Journal of Statistical Physics}\ }\textbf {\bibinfo {volume}
  {85}},\ \bibinfo {pages} {251} (\bibinfo {year} {1996})}\BibitemShut
  {NoStop}%
\bibitem [{\citenamefont {Wu}(2000)}]{Wu2000}%
  \BibitemOpen
  \bibfield  {author} {\bibinfo {author} {\bibfnamefont {C.~C.}\ \bibnamefont
  {Wu}},\ }\href {\doibase 10.1023/A:1018763008810} {\bibfield  {journal}
  {\bibinfo  {journal} {Journal of Statistical Physics}\ }\textbf {\bibinfo
  {volume} {100}},\ \bibinfo {pages} {893} (\bibinfo {year}
  {2000})}\BibitemShut {NoStop}%
\bibitem [{\citenamefont {Rietman}\ \emph {et~al.}(1992)\citenamefont
  {Rietman}, \citenamefont {Nienhuis},\ and\ \citenamefont
  {Oitmaa}}]{Rietman1992}%
  \BibitemOpen
  \bibfield  {author} {\bibinfo {author} {\bibfnamefont {R.}~\bibnamefont
  {Rietman}}, \bibinfo {author} {\bibfnamefont {B.}~\bibnamefont {Nienhuis}}, \
  and\ \bibinfo {author} {\bibfnamefont {J.}~\bibnamefont {Oitmaa}},\ }\href
  {\doibase 10.1088/0305-4470/25/24/012} {\bibfield  {journal} {\bibinfo
  {journal} {Journal of Physics A: Mathematical and General}\ }\textbf
  {\bibinfo {volume} {25}},\ \bibinfo {pages} {6577} (\bibinfo {year}
  {1992})}\BibitemShut {NoStop}%
\bibitem [{\citenamefont {Kramers}\ and\ \citenamefont
  {Wannier}(1941)}]{Krammers1941}%
  \BibitemOpen
  \bibfield  {author} {\bibinfo {author} {\bibfnamefont {H.~A.}\ \bibnamefont
  {Kramers}}\ and\ \bibinfo {author} {\bibfnamefont {G.~H.}\ \bibnamefont
  {Wannier}},\ }\href {\doibase 10.1103/PhysRev.60.252} {\bibfield  {journal}
  {\bibinfo  {journal} {Phys. Rev.}\ }\textbf {\bibinfo {volume} {60}},\
  \bibinfo {pages} {252} (\bibinfo {year} {1941})}\BibitemShut {NoStop}%
\bibitem [{\citenamefont {Jiang}\ \emph {et~al.}(2019)\citenamefont {Jiang},
  \citenamefont {Dumer}, \citenamefont {Kovalev},\ and\ \citenamefont
  {Pryadko}}]{Jiang2018}%
  \BibitemOpen
  \bibfield  {author} {\bibinfo {author} {\bibfnamefont {Y.}~\bibnamefont
  {Jiang}}, \bibinfo {author} {\bibfnamefont {I.}~\bibnamefont {Dumer}},
  \bibinfo {author} {\bibfnamefont {A.~A.}\ \bibnamefont {Kovalev}}, \ and\
  \bibinfo {author} {\bibfnamefont {L.~P.}\ \bibnamefont {Pryadko}},\ }\href
  {\doibase 10.1063/1.5039735} {\bibfield  {journal} {\bibinfo  {journal}
  {Journal of Mathematical Physics}\ }\textbf {\bibinfo {volume} {60}},\
  \bibinfo {pages} {083302} (\bibinfo {year} {2019})},\ \Eprint
  {http://arxiv.org/abs/https://doi.org/10.1063/1.5039735}
  {https://doi.org/10.1063/1.5039735} \BibitemShut {NoStop}%
\bibitem [{\citenamefont {Aizenman}(1980)}]{Aizenman1980}%
  \BibitemOpen
  \bibfield  {author} {\bibinfo {author} {\bibfnamefont {M.}~\bibnamefont
  {Aizenman}},\ }\href {https://projecteuclid.org:443/euclid.cmp/1103907767}
  {\bibfield  {journal} {\bibinfo  {journal} {Comm. Math. Phys.}\ }\textbf
  {\bibinfo {volume} {73}},\ \bibinfo {pages} {83} (\bibinfo {year}
  {1980})}\BibitemShut {NoStop}%
\bibitem [{\citenamefont {Krcmar}\ \emph {et~al.}(2008)\citenamefont {Krcmar},
  \citenamefont {Gendiar}, \citenamefont {Ueda},\ and\ \citenamefont
  {Nishino}}]{Krcmar2008}%
  \BibitemOpen
  \bibfield  {author} {\bibinfo {author} {\bibfnamefont {R.}~\bibnamefont
  {Krcmar}}, \bibinfo {author} {\bibfnamefont {A.}~\bibnamefont {Gendiar}},
  \bibinfo {author} {\bibfnamefont {K.}~\bibnamefont {Ueda}}, \ and\ \bibinfo
  {author} {\bibfnamefont {T.}~\bibnamefont {Nishino}},\ }\href {\doibase
  10.1088/1751-8113/41/12/125001} {\bibfield  {journal} {\bibinfo  {journal}
  {Journal of Physics A: Mathematical and Theoretical}\ }\textbf {\bibinfo
  {volume} {41}},\ \bibinfo {pages} {125001} (\bibinfo {year}
  {2008})}\BibitemShut {NoStop}%
\bibitem [{\citenamefont {Shima}\ and\ \citenamefont
  {Sakaniwa}(2006)}]{Shima2005}%
  \BibitemOpen
  \bibfield  {author} {\bibinfo {author} {\bibfnamefont {H.}~\bibnamefont
  {Shima}}\ and\ \bibinfo {author} {\bibfnamefont {Y.}~\bibnamefont
  {Sakaniwa}},\ }\href {\doibase 10.1088/0305-4470/39/18/010} {\bibfield
  {journal} {\bibinfo  {journal} {J. Phys.}\ }\textbf {\bibinfo {volume}
  {A39}},\ \bibinfo {pages} {4921} (\bibinfo {year} {2006})},\ \Eprint
  {http://arxiv.org/abs/cond-mat/0511539} {arXiv:cond-mat/0511539 [cond-mat]}
  \BibitemShut {NoStop}%
\bibitem [{\citenamefont {dAuriac}\ \emph {et~al.}(2001)\citenamefont
  {dAuriac}, \citenamefont {M{\'{e}}lin}, \citenamefont {Chandra},\ and\
  \citenamefont {Dou{\c{c}}ot}}]{d_Auriac2001}%
  \BibitemOpen
  \bibfield  {author} {\bibinfo {author} {\bibfnamefont {J.~C.~A.}\
  \bibnamefont {dAuriac}}, \bibinfo {author} {\bibfnamefont {R.}~\bibnamefont
  {M{\'{e}}lin}}, \bibinfo {author} {\bibfnamefont {P.}~\bibnamefont
  {Chandra}}, \ and\ \bibinfo {author} {\bibfnamefont {B.}~\bibnamefont
  {Dou{\c{c}}ot}},\ }\href {\doibase 10.1088/0305-4470/34/4/301} {\bibfield
  {journal} {\bibinfo  {journal} {Journal of Physics A: Mathematical and
  General}\ }\textbf {\bibinfo {volume} {34}},\ \bibinfo {pages} {675}
  (\bibinfo {year} {2001})}\BibitemShut {NoStop}%
\bibitem [{\citenamefont {Sausset}\ and\ \citenamefont
  {Tarjus}(2007)}]{Sausset2007}%
  \BibitemOpen
  \bibfield  {author} {\bibinfo {author} {\bibfnamefont {F.}~\bibnamefont
  {Sausset}}\ and\ \bibinfo {author} {\bibfnamefont {G.}~\bibnamefont
  {Tarjus}},\ }\href {\doibase 10.1088/1751-8113/40/43/004} {\bibfield
  {journal} {\bibinfo  {journal} {Journal of Physics A: Mathematical and
  Theoretical}\ }\textbf {\bibinfo {volume} {40}},\ \bibinfo {pages} {12873}
  (\bibinfo {year} {2007})}\BibitemShut {NoStop}%
\bibitem [{\citenamefont {Doyon}\ and\ \citenamefont
  {Fonseca}(2004)}]{Doyon2004}%
  \BibitemOpen
  \bibfield  {author} {\bibinfo {author} {\bibfnamefont {B.}~\bibnamefont
  {Doyon}}\ and\ \bibinfo {author} {\bibfnamefont {P.}~\bibnamefont
  {Fonseca}},\ }\href {\doibase 10.1088/1742-5468/2004/07/p07002} {\bibfield
  {journal} {\bibinfo  {journal} {Journal of Statistical Mechanics: Theory and
  Experiment}\ }\textbf {\bibinfo {volume} {2004}},\ \bibinfo {pages} {P07002}
  (\bibinfo {year} {2004})}\BibitemShut {NoStop}%
\bibitem [{\citenamefont {Benedetti}(2015)}]{Benedetti2015}%
  \BibitemOpen
  \bibfield  {author} {\bibinfo {author} {\bibfnamefont {D.}~\bibnamefont
  {Benedetti}},\ }\href {\doibase 10.1088/1742-5468/2015/01/p01002} {\bibfield
  {journal} {\bibinfo  {journal} {Journal of Statistical Mechanics: Theory and
  Experiment}\ }\textbf {\bibinfo {volume} {2015}},\ \bibinfo {pages} {P01002}
  (\bibinfo {year} {2015})}\BibitemShut {NoStop}%
\bibitem [{\citenamefont {Newman}\ and\ \citenamefont
  {Barkema}(1999)}]{Newman1999}%
  \BibitemOpen
  \bibfield  {author} {\bibinfo {author} {\bibfnamefont {M.}~\bibnamefont
  {Newman}}\ and\ \bibinfo {author} {\bibfnamefont {G.}~\bibnamefont
  {Barkema}},\ }\href {https://books.google.de/books?id=J5aLdDN4uFwC} {\emph
  {\bibinfo {title} {Monte Carlo Methods in Statistical Physics}}}\ (\bibinfo
  {publisher} {Clarendon Press},\ \bibinfo {year} {1999})\BibitemShut {NoStop}%
\bibitem [{\citenamefont {Landau}\ and\ \citenamefont
  {Binder}(2009)}]{Landau2009}%
  \BibitemOpen
  \bibfield  {author} {\bibinfo {author} {\bibfnamefont {D.}~\bibnamefont
  {Landau}}\ and\ \bibinfo {author} {\bibfnamefont {K.}~\bibnamefont
  {Binder}},\ }\href {https://books.google.de/books?id=hrIhAwAAQBAJ} {\emph
  {\bibinfo {title} {A Guide to Monte Carlo Simulations in Statistical
  Physics}}}\ (\bibinfo  {publisher} {Cambridge University Press},\ \bibinfo
  {year} {2009})\BibitemShut {NoStop}%
\bibitem [{\citenamefont {Botet}\ \emph {et~al.}(1982)\citenamefont {Botet},
  \citenamefont {Jullien},\ and\ \citenamefont {Pfeuty}}]{Botet1982}%
  \BibitemOpen
  \bibfield  {author} {\bibinfo {author} {\bibfnamefont {R.}~\bibnamefont
  {Botet}}, \bibinfo {author} {\bibfnamefont {R.}~\bibnamefont {Jullien}}, \
  and\ \bibinfo {author} {\bibfnamefont {P.}~\bibnamefont {Pfeuty}},\ }\href
  {\doibase 10.1103/PhysRevLett.49.478} {\bibfield  {journal} {\bibinfo
  {journal} {Phys. Rev. Lett.}\ }\textbf {\bibinfo {volume} {49}},\ \bibinfo
  {pages} {478} (\bibinfo {year} {1982})}\BibitemShut {NoStop}%
\bibitem [{\citenamefont {Singh}\ and\ \citenamefont
  {Chakravarty}(1987{\natexlab{a}})}]{singh87}%
  \BibitemOpen
  \bibfield  {author} {\bibinfo {author} {\bibfnamefont {R.~R.~P.}\
  \bibnamefont {Singh}}\ and\ \bibinfo {author} {\bibfnamefont
  {S.}~\bibnamefont {Chakravarty}},\ }\href {\doibase 10.1103/PhysRevB.36.546}
  {\bibfield  {journal} {\bibinfo  {journal} {Phys. Rev. B}\ }\textbf {\bibinfo
  {volume} {36}},\ \bibinfo {pages} {546} (\bibinfo {year}
  {1987}{\natexlab{a}})}\BibitemShut {NoStop}%
\bibitem [{\citenamefont {Singh}\ and\ \citenamefont
  {Chakravarty}(1987{\natexlab{b}})}]{singh2}%
  \BibitemOpen
  \bibfield  {author} {\bibinfo {author} {\bibfnamefont {R.~R.~P.}\
  \bibnamefont {Singh}}\ and\ \bibinfo {author} {\bibfnamefont
  {S.}~\bibnamefont {Chakravarty}},\ }\href {\doibase 10.1103/PhysRevB.36.559}
  {\bibfield  {journal} {\bibinfo  {journal} {Phys. Rev. B}\ }\textbf {\bibinfo
  {volume} {36}},\ \bibinfo {pages} {559} (\bibinfo {year}
  {1987}{\natexlab{b}})}\BibitemShut {NoStop}%
\bibitem [{\citenamefont {Oitmaa}\ \emph {et~al.}(2006)\citenamefont {Oitmaa},
  \citenamefont {Hamer}, \citenamefont {Zheng},\ and\ \citenamefont
  {Press}}]{oitmaa2006}%
  \BibitemOpen
  \bibfield  {author} {\bibinfo {author} {\bibfnamefont {J.}~\bibnamefont
  {Oitmaa}}, \bibinfo {author} {\bibfnamefont {C.}~\bibnamefont {Hamer}},
  \bibinfo {author} {\bibfnamefont {W.}~\bibnamefont {Zheng}}, \ and\ \bibinfo
  {author} {\bibfnamefont {C.~U.}\ \bibnamefont {Press}},\ }\href
  {https://books.google.de/books?id=VCYUMPDK73IC} {\emph {\bibinfo {title}
  {Series Expansion Methods for Strongly Interacting Lattice Models}}}\
  (\bibinfo  {publisher} {Cambridge University Press},\ \bibinfo {year}
  {2006})\BibitemShut {NoStop}%
\bibitem [{\citenamefont {Chaikin}\ and\ \citenamefont
  {Lubensky}(2000)}]{Chaikin2000}%
  \BibitemOpen
  \bibfield  {author} {\bibinfo {author} {\bibfnamefont {P.~M.}\ \bibnamefont
  {Chaikin}}\ and\ \bibinfo {author} {\bibfnamefont {T.~C.}\ \bibnamefont
  {Lubensky}},\ }\href {https://books.google.de/books?id=P9YjNjzr9OIC} {\emph
  {\bibinfo {title} {{Principles of Condensed Matter Physics}}}}\ (\bibinfo
  {publisher} {Cambridge University Press},\ \bibinfo {year}
  {2000})\BibitemShut {NoStop}%
\bibitem [{\citenamefont {Baxter}(2013)}]{Baxter2013}%
  \BibitemOpen
  \bibfield  {author} {\bibinfo {author} {\bibfnamefont {R.}~\bibnamefont
  {Baxter}},\ }\href {https://books.google.de/books?id=eQzCAgAAQBAJ} {\emph
  {\bibinfo {title} {Exactly Solved Models in Statistical Mechanics}}},\ Dover
  Books on Physics\ (\bibinfo  {publisher} {Dover Publications},\ \bibinfo
  {year} {2013})\BibitemShut {NoStop}%
\bibitem [{\citenamefont {Madras}\ and\ \citenamefont {Wu}(2005)}]{hypSAW}%
  \BibitemOpen
  \bibfield  {author} {\bibinfo {author} {\bibfnamefont {N.}~\bibnamefont
  {Madras}}\ and\ \bibinfo {author} {\bibfnamefont {C.~C.}\ \bibnamefont
  {Wu}},\ }\href@noop {} {\bibfield  {journal} {\bibinfo  {journal}
  {Combinatorics, Probability and Computing}\ }\textbf {\bibinfo {volume}
  {14}},\ \bibinfo {pages} {523} (\bibinfo {year} {2005})}\BibitemShut
  {NoStop}%
\bibitem [{\citenamefont {Kumar}(1976)}]{kumar1976two}%
  \BibitemOpen
  \bibfield  {author} {\bibinfo {author} {\bibfnamefont {D.}~\bibnamefont
  {Kumar}},\ }\href@noop {} {\bibfield  {journal} {\bibinfo  {journal}
  {Pramana}\ }\textbf {\bibinfo {volume} {7}},\ \bibinfo {pages} {28} (\bibinfo
  {year} {1976})}\BibitemShut {NoStop}%
\bibitem [{\citenamefont {J.~Kollar}\ \emph {et~al.}(2019)\citenamefont
  {J.~Kollar}, \citenamefont {Fitzpatrick},\ and\ \citenamefont
  {A.~Houck}}]{Kollar2019}%
  \BibitemOpen
  \bibfield  {author} {\bibinfo {author} {\bibfnamefont {A.}~\bibnamefont
  {J.~Kollar}}, \bibinfo {author} {\bibfnamefont {M.}~\bibnamefont
  {Fitzpatrick}}, \ and\ \bibinfo {author} {\bibfnamefont {A.}~\bibnamefont
  {A.~Houck}},\ }\href {\doibase 10.1038/s41586-019-1348-3} {\bibfield
  {journal} {\bibinfo  {journal} {Nature}\ }\textbf {\bibinfo {volume} {571}}
  (\bibinfo {year} {2019}),\ 10.1038/s41586-019-1348-3}\BibitemShut {NoStop}%
\bibitem [{\citenamefont {Dennis}\ \emph {et~al.}(2002)\citenamefont {Dennis},
  \citenamefont {Kitaev}, \citenamefont {Landahl},\ and\ \citenamefont
  {Preskill}}]{dklp}%
  \BibitemOpen
  \bibfield  {author} {\bibinfo {author} {\bibfnamefont {E.}~\bibnamefont
  {Dennis}}, \bibinfo {author} {\bibfnamefont {A.}~\bibnamefont {Kitaev}},
  \bibinfo {author} {\bibfnamefont {A.}~\bibnamefont {Landahl}}, \ and\
  \bibinfo {author} {\bibfnamefont {J.}~\bibnamefont {Preskill}},\ }\href@noop
  {} {\bibfield  {journal} {\bibinfo  {journal} {Journal of Mathematical
  Physics}\ }\textbf {\bibinfo {volume} {43}},\ \bibinfo {pages} {4452}
  (\bibinfo {year} {2002})}\BibitemShut {NoStop}%
\bibitem [{\citenamefont {Moran}(1997)}]{moran}%
  \BibitemOpen
  \bibfield  {author} {\bibinfo {author} {\bibfnamefont {J.~F.}\ \bibnamefont
  {Moran}},\ }\href {\doibase http://dx.doi.org/10.1016/S0012-365X(96)00102-1}
  {\bibfield  {journal} {\bibinfo  {journal} {Discrete Mathematics}\ }\textbf
  {\bibinfo {volume} {173}},\ \bibinfo {pages} {151 } (\bibinfo {year}
  {1997})}\BibitemShut {NoStop}%
\bibitem [{\citenamefont {Hasegawa}\ \emph {et~al.}(2007)\citenamefont
  {Hasegawa}, \citenamefont {Sakaniwa},\ and\ \citenamefont
  {Shima}}]{hasegawa}%
  \BibitemOpen
  \bibfield  {author} {\bibinfo {author} {\bibfnamefont {I.}~\bibnamefont
  {Hasegawa}}, \bibinfo {author} {\bibfnamefont {Y.}~\bibnamefont {Sakaniwa}},
  \ and\ \bibinfo {author} {\bibfnamefont {H.}~\bibnamefont {Shima}},\ }\href
  {\doibase https://doi.org/10.1016/j.susc.2007.04.207} {\bibfield  {journal}
  {\bibinfo  {journal} {Surface Science}\ }\textbf {\bibinfo {volume} {601}},\
  \bibinfo {pages} {5232 } (\bibinfo {year} {2007})},\ \bibinfo {note}
  {proceedings of the 10th ISSP International Symposium on Nanoscience at
  Surfaces}\BibitemShut {NoStop}%
\bibitem [{\citenamefont {Breuckmann}\ and\ \citenamefont
  {Terhal}(2016)}]{hyperbolic_constr_thresh}%
  \BibitemOpen
  \bibfield  {author} {\bibinfo {author} {\bibfnamefont {N.~P.}\ \bibnamefont
  {Breuckmann}}\ and\ \bibinfo {author} {\bibfnamefont {B.~M.}\ \bibnamefont
  {Terhal}},\ }\href@noop {} {\bibfield  {journal} {\bibinfo  {journal} {IEEE
  Transactions on Information Theory}\ }\textbf {\bibinfo {volume} {62}},\
  \bibinfo {pages} {3731} (\bibinfo {year} {2016})}\BibitemShut {NoStop}%
\end{thebibliography}%
\end{document}